\documentclass[twocolumn,showpacs,preprintnumbers,amsmath,amssymb,prl,floatfix]{revtex4}

\usepackage{graphicx}
\usepackage{dcolumn}
\usepackage{bm}
\usepackage{color}
\usepackage{mathdots}

\begin{document}

%
%

\title{Quantum Phases of Two-Component Bosons\\ with Spin-Orbit Coupling in Optical Lattices
}

\author{Daisuke Yamamoto$^1$, {I. B. Spielman}$^2$, {C. A. R. S\'a de Melo}$^{2,3}$}
\affiliation{$^1$Department of Physics and Mathematics, Aoyama-Gakuin University, Sagamihara, Kanagawa 252-5258, Japan}
\affiliation{$^2$Joint Quantum Institute, National Institute of Standards and Technology, 
and University of Maryland, Gaithersburg, Maryland 20899, USA}
\affiliation{$^3$School of Physics, Georgia Institute of Technology, Atlanta, Georgia 30332, USA}
\date{\today}
\begin{abstract}
{Ultracold bosons in optical lattices are one of the few systems where bosonic matter is known to exhibit strong correlations. {Here we push the frontier of our understanding of interacting bosons in optical lattices by adding} synthetic spin-orbit coupling, and show that new kinds of density- and chiral-orders develop. {The competition between the optical lattice period and the spin-orbit coupling length -- which can be made comparable in experiments -- along with} the spin hybridization induced {by a transverse field (i.e., Rabi coupling)} and interparticle interactions create a rich variety of quantum phases including {uniform, non-uniform and phase-separated superfluids, as well as} Mott insulators. 
The spontaneous symmetry breaking phenomena at the transitions between them are explained by a two-order-parameter Ginzburg-Landau model with multiparticle umklapp processes. Finally, in order to characterize each phase, we calculated {their experimentally measurable crystal momentum distributions}. }
\end{abstract}
\maketitle

{The physics of spin-orbit coupling (SOC), which links the spin and momentum degrees of freedom in quantum particles, is ubiquitous in nature, ranging from the microscopic world of atoms, such as Hydrogen, to macroscopic solid materials, such as semiconductors.} Recently, the effects of {SOC} have been explored in condensed matter physics in connection with topological insulators~\cite{hasan-10}, as well as with topological superconductors~\cite{qi-11}, and superconductors without inversion symmetry~\cite{bauer-12}. 
In these naturally occurring systems, it is very difficult to control the magnitude of {SOC} {and yet more difficult to study correlated bosons. However it is now possible} to create controllable artificial {SOC for trapped ultracold} fermionic and bosonic atoms~\cite{lin-11,zhang-12,wang-12,cheuk-12,reviews,li-17}, {the physics of which was recently} analyzed theoretically in 
the continuum limit~\cite{lin-11,wu-11,ho-11,li-12,li-13}. One of the emerging frontiers in this broad area of physics {is the} interplay of the spin-orbit and lattice characteristic lengths, which can be made comparable in optical lattice {systems, where additional contributions from a Zeeman field and strong local interactions also play an important role.}

{In this Letter, we obtain first the ground-state phase diagrams}  for two-component $(\uparrow, \downarrow)$ bosons
in the presence of artificial {SOC}, {an effective Zeeman field (created from Rabi coupling and detuning),} and local interactions. With zero detuning, we identify four phases: {uniform, non-uniform and phase-separated superfluids, along with Mott insulating phases}, depending on interactions. {Secondly,} we develop a Ginzburg-Landau theory for further characterizing these phases. {Lastly,} we calculate their crystal momentum distributions, which can be compared with experiments.

{To describe the quantum phases of two-component bosons with {SOC}, we begin by introducing the independent particle Hamiltonian
\begin{eqnarray}
\hat{\mathcal{H}}_0
=
\sum_{\bm{k}}\left(\begin{array}{cc}
\hat{b}_{\bm{k}\uparrow}^\dagger  &  
\hat{b}_{\bm{k}\downarrow}^\dagger\end{array}\right) \left(\begin{array}{cc}
\epsilon_{\bm{k}\uparrow}-\mu &  {\hbar \Omega/2}\\
{\hbar \Omega/2}&\epsilon_{\bm{k}\downarrow}-\mu\end{array}\right) \left(\begin{array}{c}
\hat{b}_{\bm{k}\uparrow}  \\  \hat{b}_{\bm{k}\downarrow}\end{array}\right)
\end{eqnarray}
in momentum space. Here, {$\epsilon_{\bm{k}s}=-2t[\cos (k_x+s{k_T})+\cos k_y+\cos k_z]+s{\hbar \delta/2}$} for a three-dimensional (3D) optical lattice and $\bm{k}_T=(k_T,0,0)$ is {the SOC momentum}. {The} length scale $2\pi/k_T$ is of the order of the  optical lattice spacing $a$, chosen to be one. The operator $\hat{b}_{\bm{k}s}^\dagger$ describes a creation of {$s \in \left\{\uparrow,\downarrow\right\} \equiv \left\{+,-\right\}$ boson with momentum $\bm{k}$. In addition,} the chemical potential $\mu$ tunes the average particle density $\rho=\rho_\uparrow+\rho_\downarrow\equiv \sum_{\bm{k}s}\langle\hat{b}_{\bm{k}s}^\dagger\hat{b}_{\bm{k}s}\rangle/M$ with $M$ being the number of lattice sites. 
In cold-atom experiments, the effective Zeeman {energy ${\boldsymbol \Omega}\cdot\hat{{\bf F}}$ with ${\boldsymbol \Omega}=(\Omega,0,\delta)$ and $\hat{{\bf F}}$ being the total angular momentum operator for spin-1/2} has two parts: spin flips through the Rabi frequency $\Omega$ and a Zeeman shift via the detuning $\delta$. }The Hamiltonian above can be engineered in the laboratory either through
Raman processes~\cite{lin-11,zhang-12,higbie-02} or via radio-frequency chips~\cite{goldman-10,anderson-13}.

The diagonalization of $\hat{\mathcal{H}}_0$ gives two energy branches
\begin{eqnarray*}
{E_{\bm{k}\pm}
=
\left(\epsilon_{\bm{k}\uparrow}
+
\epsilon_{\bm{k}\downarrow}
-2\mu\pm\sqrt{\left(\epsilon_{\bm{k}\uparrow}
-\epsilon_{\bm{k}\downarrow}\right)^2+(\hbar \Omega)^2}\right)/2.}
\end{eqnarray*}
For $\delta=0$ and small {$\hbar\Omega/t$}, the lower branch $E_{\bm{k}-}$ has two degenerate minima at $k_x \approx \pm k_T$ and $k_y=k_z=0$. 
The two minima {approach as} the Rabi frequency 
(spin-hybridization) $\Omega$ is increased, and eventualy they collapse 
into a single minimum at ${\bm k}={\bm 0}$ when {$\hbar\Omega/t\geq 4\sin k_T \tan k_T$}. This double-minimum structure, the 
introduction of a new length scale $1/k_T$ and 
the interactions between particles
\begin{eqnarray}
\hat{\mathcal{H}}_{\rm int}
=
\frac{1}{2}\sum_{\bm{k}\bm{q}}\sum_{ss^\prime}U_{ss^\prime}\hat{b}_{\bm{k}s}^\dagger \hat{b}_{\bm{k}+\bm{q}s^\prime}^\dagger \hat{b}_{\bm{k}-\bm{q}s^\prime} \hat{b}_{\bm{k}s}
\label{hamiltonian_interactions}
\end{eqnarray}
provide additional contributions that are absent in the standard spinless Bose-Hubbard system~\cite{greiner-02}. In this work, {we explore the special case} where the same spin repulsions $U_{ss}$ are nearly identical ($U_{\uparrow\uparrow} \approx U_{\downarrow\downarrow} = U$), but the opposite spin repulsion is different from $U$, that is, $U\geq U_{\uparrow\downarrow}=U_{\downarrow\uparrow}\geq 0$. 
For instance, in the case of a mixture of the $m_F=0$ ($\downarrow$) and $m_F=-1$ ($\uparrow$) states from the $F=1$ manifold of $^{87}$Rb, these repulsions are nearly identical ($U_{\uparrow\uparrow} \approx U_{\downarrow\downarrow} \approx U_{\uparrow\downarrow}$)~\cite{widera-06}.

We begin our analysis of the quantum phases of this complex system by 
investigating first the regime of weak repulsive interactions. 
{In the semi-classical regime ($U\ll t\rho$), the bosonic fields $\hat{b}_{\bm{k}s}$ can be written as
%
$
\hat{b}_{\bm{k}s}
=
\sum^\prime_{q}\sqrt{M}\psi_{qs}\delta_{{\bm k}=(q,0,0)}
+
\hat{a}_{\bm{k}s},
$
%
where $\sqrt{M}\psi_{qs}$ and $\hat{a}_{\bm{k}s}$ describe the Bose-Einstein condensate (BEC) with momentum ${\bm k}=(q,0,0)$ and the residual bosons outside the condensate, respectively.} Considering the single and double minima features of $E_{\bm{k}-}$ within the first Brillouin zone, we allow for multiple BECs with different momenta and take the sum $\sum^\prime_{q}$ to be
over the set of possible momenta $\{ q \}$ along the $(k_x,0,0)$ direction. 
The energy per site of the condensates is 
\begin{eqnarray}
\frac{E_0}{M}
=
\sum^\prime_{q}\left(\begin{array}{cc}
\psi_{q\uparrow}^\ast  &  \psi_{q\downarrow}^\ast\end{array}\right) 
\left(\begin{array}{cc}
\epsilon_{\bm{k}\uparrow}-\mu &  {\hbar\Omega/2}\\
{\hbar\Omega/2}&\epsilon_{\bm{k}\downarrow}-\mu\end{array}\right) 
\left(
\begin{array}{c}
\psi_{q\uparrow}  \\  \psi_{q\downarrow}\end{array}
\right) + \nonumber
\\ 
\sum^\prime_{\{q_i\}} 
\left[\frac{U}{2}
\sum_s \psi_{q_1s}^\ast\psi_{q_2s}^\ast\psi_{q_3s}\psi_{q_4s}
+
U_{\uparrow\downarrow} 
\psi_{q_1\uparrow}^\ast\psi_{q_2\downarrow}^\ast
\psi_{q_3\downarrow}\psi_{q_4\uparrow}\right],
\label{GP_energy}
\end{eqnarray}
where the sum $\sum^\prime_{\{q_i\}}$ is over momenta $q_i$ satisfying
momentum conservation ${q_1 + q_2 = q_3 + q_4}$ 
$\left[ {\rm mod}~2\pi \right]$.

\begin{figure}[t]
\includegraphics[scale=0.4]{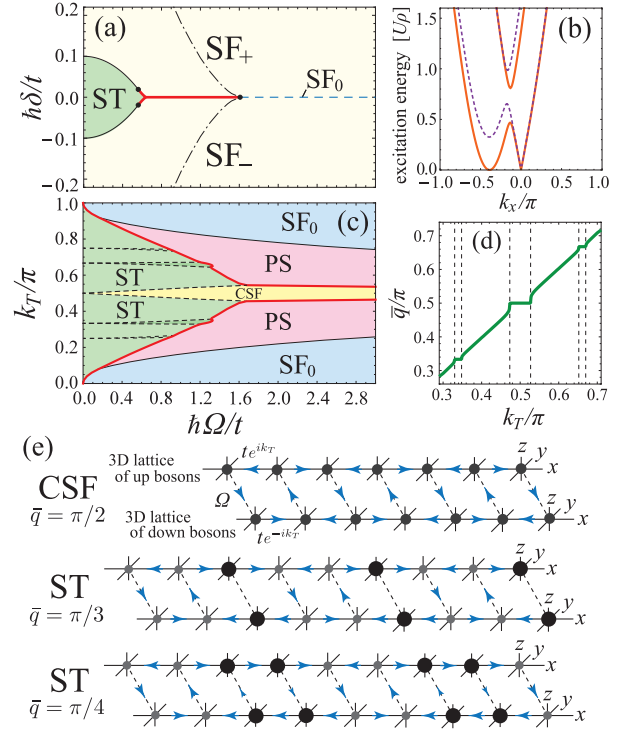}
\caption{\label{PD_GP_Omega_Delta}
(color online). Ground-state properties in the weak-coupling regime 
with $U/t=1/\rho$ and $U_{\uparrow\downarrow}=0.9U$. 
(a) Phase diagram of detuning {$\delta$} versus Rabi frequency {$\Omega$} for $k_T=0.2\pi$. The thick red (thin black) curves denote first- (second-) 
order transitions and the black dots indicate multicritical points. {In the $\delta>0$ ($\delta<0$) region to the left side of the dash-dotted line, the SF$_-$ (SF$_+$) exists only as a metastable state.} (b) Roton-like softening in the 
elementary excitations for quasimomentum $\bm{k}=(k_x,0,0)$ 
and {$\hbar \Omega/t=0.4$}. We set {$\hbar\delta/t=0.4$} (in SF$_+$) 
for the dotted lines and {$\hbar\delta/t=0.06$} (at the SF$_+$-ST boundary) 
for the solid lines. (c) The $k_T$ dependence of the ground state 
when $\rho_\uparrow=\rho_\downarrow$ {($\delta=0$)}. {The yellow and darker green regions limited by the black-dashed and red lines are the CSF and period-locked ST phases illustrated in (e). }
(d) The plateaux in $\bar{q}$ of the CSF and {period-locked} ST phases as a function of $k_T$ for {$\hbar\Omega/t=1.0$}. The dashed lines denote the width of dominant plateaux with commensurate wavenumber $\bar{q}$. (e) Density (the size of dots) and chiral (the direction of arrows) patterns in the commensurate phases. }
\end{figure}

After minimization of Eq.~(\ref{GP_energy}) with respect to $\psi_{qs}$ 
and $\{q\}$, we find four different ground states
as shown in Fig.~\ref{PD_GP_Omega_Delta}(a) for {the weak-coupling regime with} parameters 
$U=t/\rho$, $U_{\uparrow\downarrow}=0.9U$, and $k_T=0.2\pi$. 
In the superfluid phases (SF$_\pm$), the set of BEC momenta $\{q\}$ 
consists of a single value ($\bar{q}>0$ in SF$_+$ and $-\bar{q}<0$ in SF$_-$) 
since the detuning $\delta$ tilts the single-particle spectrum and lifts 
the degeneracy of the double minima in $E_{{\bf k}-}$. 
In these ``single-$q$'' states, the particle density is uniform, while 
the phase of the condensate spatially varies with pitch vector 
$(\pm\bar{q},0,0)$. 
In the striped superfluid (ST) phase for relatively small {$\hbar\Omega/t$}, a BEC is formed with 
two different momenta $-\bar{q}_1$ and $\bar{q}_2$ due to a double-minimum dispersion in $E_{{\bf k}-}$. The interference of these two momenta leads to a non-uniform density profile along the $x$ direction, resulting in a stripe pattern. {Moreover, the scattering process under momentum conservation ${q_1 + q_2 = q_3 + q_4}$ with $q_3=q_4=-\bar{q}_1$ and $q_2=\bar{q}_2$  (or vice-versa) gives rise to a higher harmonic component with $q_1=-2\bar{q}_1-\bar{q}_2$ (or $q_1=\bar{q}_1+2\bar{q}_2$). Similar processes generate higher harmonics with interval $\bar{q}_1+\bar{q}_2$, thus making the set $\{q\}$ have a large number of different momenta $-\bar{q}_1+n(\bar{q}_1+\bar{q}_2)$, where $n$ is an integer.

When {$\hbar\Omega/t$} is large, 
the SF$_+$ and SF$_-$ phases are continuously connected at $\delta = 0$ through the conventional superfluid (SF$_0$) with zero-momentum BEC. However, when {$\hbar\Omega/t$} has intermediate values, a direct first-order transition from SF$_+$ to SF$_-$ takes place, and thus the spin population difference $\rho_\downarrow-\rho_\uparrow$ exhibits a sudden jump from positive 
to negative. 
Therefore, in the experimental situation 
where the population of each spin is balanced 
($\rho_\downarrow$=$\rho_\uparrow$), the system is unstable against 
spatial phase separation (PS) of spin-down-rich SF$_+$ 
and spin-up-rich SF$_-$ states.

The quadratic part of the Hamiltonian in terms of $\hat{a}_{\bm{k}s}$, 
$
\hat {\mathcal H}_{B} =
\sum_{\bm{k}}\hat{\bm{a}}^\dagger_{\bm{k}} 
H^{(2)}_{\bm{k}}\hat{\bm{a}}_{\bm{k}}
$, 
is a generalized Bogoliubov Hamiltonian
and includes quantum fluctuations outside the condensate perturbatively. 
We diagonalize $\hat {\mathcal H}_{B}$ numerically via a generalized
Bogoliubov transformation~\cite{colpa-78,supplemental-material}, and obtain the spectrum of elementary excitations. 
In Fig.~\ref{PD_GP_Omega_Delta}(b) we show 
typical excitation spectra of the SF$_+$ states. We can see a 
roton-like minimum at a finite quasimomentum with the excitation energy 
approaching zero as the detuning $\delta$ is decreased (increased) as we move 
from the SF$_+$ (SF$_-$) phase towards the ST phase. The transition from 
SF$_+$ (or SF$_-$) to ST is induced by the softening of the 
roton-like minimum, similar to the standard superfluid-supersolid 
transition~\cite{ji-15}. The momentum of the roton-like excitations {largely} determines the 
characteristic reciprocal vector $\bar{q}_1+\bar{q}_2$ of the ST state 
resulting from the phase transition. {Furthermore, the transition from SF$_+$ or SF$_-$ to the ST phase can also be first order as indicated by the red solid line shown in Fig.~\ref{PD_GP_Omega_Delta}(a).} In this case, the energy gap 
of roton-like excitations jumps discontinuously to zero at
the SF$_\pm$/ST boundary.

The weak coupling phase diagram shown in Fig.~\ref{PD_GP_Omega_Delta}(a) reveals ground states which are very similar to those
in the continuum limit~\cite{lin-11,wu-11,ho-11,li-12,li-13}, where the band structure 
due to the optical lattice is not important. 
However, the phase diagram of 
{SOC momentum} $k_T/\pi$ versus {$\hbar\Omega/t$} at $\rho_\downarrow$=$\rho_\uparrow$, 
shown in Fig.~\ref{PD_GP_Omega_Delta}(c), illustrates the remarkable 
competition between the intrinsic reciprocal vector of the underlying 
optical lattice and characteristic vector $\bar{q}_1+\bar{q}_2$ of the ST phase. In the spin symmetric case {($\rho_\uparrow=\rho_\downarrow$), the two wavevectors $\bar{q}_1$ and $\bar{q}_2$ are equal, that is, $\bar{q}_1=\bar{q}_2\equiv \bar{q}$ leading to $\bar{q}_1 + \bar{q}_2 = 2\bar{q}$.} The phase diagram of 
$k_T/\pi$ versus {$\hbar\Omega/t$} in the range of $k_T=\pi$ to $2\pi$ 
is exactly the same as that of Fig.~\ref{PD_GP_Omega_Delta}(c) 
since the lattice Hamiltonian 
$\hat{\mathcal{H}}_0+\hat{\mathcal{H}}_{\rm int}$ 
is invariant under the gauge tranformation 
$\hat{b}_{{\bm k}s}\rightarrow \hat{b}_{{\bm k}+(\pi,0,0)s}${, as easily verified by direct substitution}.

In Fig.~\ref{PD_GP_Omega_Delta}{(d)}, when $k_T$ is nearly commensurate to the lattice reciprocal wavenumber $2\pi$, such as $k_T\approx \pi/4$, $2\pi/3$, and $\pi/2$, 
the pitch vector $\bar{q}$ of the ST state spontaneously takes 
an exact commensurate value over a finite range of $k_T$. 
As a result, the curve of $\bar{q}/\pi$ versus $k_T/\pi$ exhibits multiple plateaux in the ST phase. 
This effect can be attributed to umklapp processes 
$q_1+q_2-q_3-q_4=2\pi n$ with 
nonzero integer $n$ that contribute 
to lower the energy of the system. 
{In particular, when $k_T\approx \pi/2$, BEC occurs with only two momenta $\pm\bar{q}=\pm \pi/2$ since all the higher-harmonics momenta are reduced to $\pm \pi/2$ due to the Brillouin zone periodicity. In this special case{ where $\bar{q}/\pi=1/2$, the interference of the two momenta does not lead to striped density pattern, but to $Z_2$ chiral symmetry breaking.} This state is analogous to the {chiral superfluid (CSF)} state, which has been discussed in Bose-Hubbard ladders~\cite{orignac-01,dhar-12,atala-14,greschner-arXiv}. In the present case, the 3D lattices for the two spin components and the Rabi couplings play the role of rails and rungs, respectively, of a synthetic ``two-leg ladder'' in four (three spatial plus one extra spin) dimensions as illustrated in Fig.~\ref{PD_GP_Omega_Delta}(e). For other commensurate {ST phases, where $\bar{q}/\pi$ takes an irreducible fraction $\zeta/\eta$ with $\zeta$ and $\eta$ being integers}, the superfluid phases break $Z_\eta$ symmetry, but preserve a stripe pattern in the atom density. The stabilization of these commensurate phases is} a specific feature of 
spin-orbit coupled systems in optical lattices {with interactions} {and are completely absent} in {interacting} continuum systems. Had we illustrated all the possible commensurate/incommensurate transitions in Fig.~\ref{PD_GP_Omega_Delta}(d), 
the graph of ${\bar q}/\pi$ versus $k_T/\pi$ would {have had an infinite number of steps at rational values of $\bar{q}/\pi$, producing a mathematical function known as the Devil's staircase. }

\begin{figure}[t]
\includegraphics[scale=0.46]{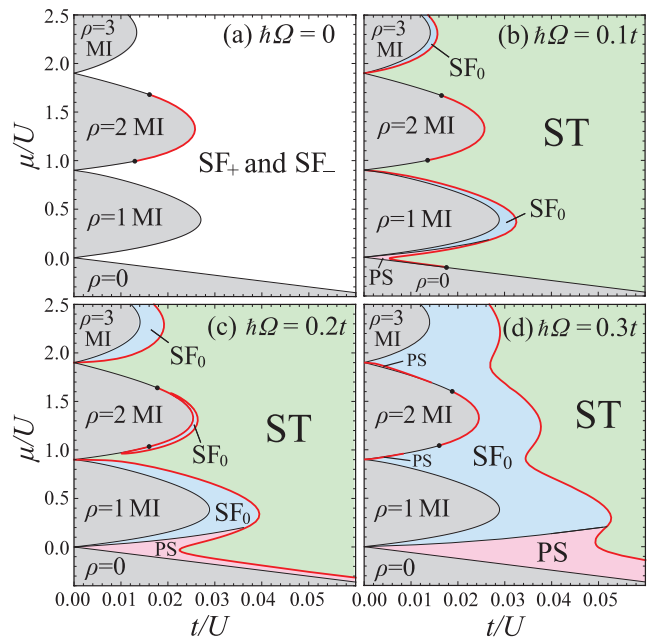}
\caption{\label{PD_t_mu}
(color online). Ground-state phase diagrams in the ($t/U$,$\mu/U$) plane, obtained by the Gutzwiller self-consistent calculations for different values of {$\hbar\Omega/t$}. We set the other parameters as $U_{\uparrow\downarrow}=0.9U$, $k_T=0.2\pi$, and $\rho_\uparrow=\rho_\downarrow$ {($\delta=0$)}. }
\end{figure}
In addition to the interplay between different length/momentum scales 
discussed above in the weak coupling regime, another particular feature
of lattice systems is the existence of Mott insulator (MI) phases induced 
by strong interactions and commensurate particle fillings. 
To describe the Mott physics in the presence of {SOC} 
and Zeeman fields, 
we employ the Gutzwiller variational method~\cite{supplemental-material}. Under the assumption that the ground state is given by a direct product state in real space, the Hamiltonian can be mapped into an effective single-site problem with variational mean fields $\psi_{is}\equiv\langle \hat{b}_{is} \rangle$, where $\hat{b}_{is}$ is the annihilation operator of spin-$s$ boson at lattice site $i$. 
To deal with the ST phases, we solve simultaneously all inequivalent single-site problems connected via mean fields due to the nonuniformity. Here, we consider up to $2\times 10^3$ mean fields $\psi_{is}$ along the $x$ direction for each spin and thus the momentum resolution is $\delta k_x\sim 0.001\pi$~\cite{supplemental-material}, while the $y$ and $z$ directions are assumed to be uniform.
{In the ST phase, the inhomogeneous state is a result of the length scale introduced by the SOC, while in the absence of SOC a new length scale leading to a supersolid state appears due to long-range interactions~\cite{scarola-06}.}

{Figure~\ref{PD_t_mu} shows phase diagrams in the $\mu/U$-$t/U$ plane} for several values of the Rabi frequency $\Omega$ in the spin symmetric case 
$\rho_\uparrow=\rho_\downarrow$ ($\delta = 0$). 
In Fig.~\ref{PD_t_mu}(a), where there is no hybridization of the two spin components ({$\hbar \Omega/t=0$}), the phase boundaries of the MI lobes are identical 
to those in the absence of {SOC}~\cite{yamamoto-13} 
since the gauge transformation 
$\hat{b}_{{\bm k}s}\rightarrow \hat{b}_{{\bm k}+s{\bm k_T}s}$ eliminates 
the momentum transfer ${\bm k_T}$ from the problem. 
{The even-filling Mott transitions become first order in a two-component Bose-Hubbard model} for large inter-component repulsions 
(for example, $U_{\uparrow\downarrow}\gtrsim 0.68U$ 
when $\rho=2$)~\cite{yamamoto-13,kuklov-04,chen-10,ozaki-12}. 
In the superfluid phase outside the Mott lobes for $k_T\neq 0$, 
the spin-down and spin-up bosons independently form 
the SF$_+$ state with $\bar{q}=k_T$ and SF$_-$ with $-\bar{q}=-k_T$, 
respectively.

The phase diagrams displayed in Figs.~\ref{PD_t_mu}(b-d), illustrate the 
effects of increasing {$\hbar\Omega$}. 
When the Rabi frequency $\Omega$ is non-vanishing, the two spin 
components {mix}, forming a nonuniform ST state with two opposite 
momenta $-\bar{q}$ and $\bar{q}$ and their associated higher harmonics{, analogous to the stripe phase in continuum systems.} 
{Figure~\ref{PD_t_mu}(b) shows} that the 
transition from the odd-filling MI to the ST phase occurs via an intermediate 
SF$_0$ state. A direct transition to 
the ST state occurs only for very small {$\hbar\Omega/t$} 
(not shown: {$\hbar\Omega/t\lesssim 0.04$} for $\rho=1$). 
As seen in Figs.~\ref{PD_t_mu}(c-d), when the value of {$\hbar\Omega/t$} 
is increased, the SF$_0$ phase also emerges near the tip of 
the $\rho=2$ MI lobe, and eventually joins other SF$_0$ regions.
{The SF$_+$ and SF$_-$ states only phase separate} for small 
fillings $\rho\lesssim 1$ and a very narrow region around 
the $\rho=2$ MI lobe for large {$\hbar\Omega/t$}.

\begin{figure}[t]
\includegraphics[scale=0.38]{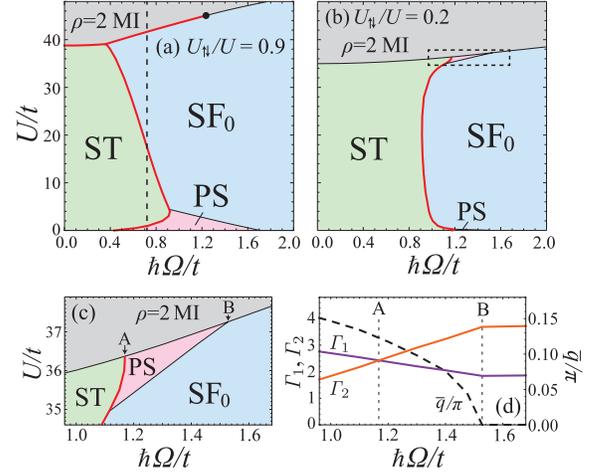}
\caption{\label{PD_GW_Omega_U}
(color online). Nonuniform superfluid-insulator transitions 
at $\rho=2$ for (a) $U_{\uparrow\downarrow}=0.9U$ and 
(b) $U_{\uparrow\downarrow}=0.2U$. We set $k_T=0.2\pi$ and 
$\rho_\uparrow=\rho_\downarrow$ {($\delta=0$)}. The vertical dashed line in 
(a) marks {$\hbar\Omega/t=0.72$}. The enlarged view of the region 
indicated by the dashed box in (b) is shown in (c). 
The fourth-order Ginzburg-Landau coefficients and the value of $\bar{q}$ along the MI transition line of (c) are plotted in (d). }
\end{figure}

To see the interplay between local correlations
and spin mixing, we plot in Figs.~\ref{PD_GW_Omega_U}(a-c) phase
diagrams of $U/t$ versus {$\hbar\Omega/t$} for fixed density $\rho=2$. 
As shown in Fig.~\ref{PD_GW_Omega_U}(a), large spin hybridization 
$\Omega$ mixes the two spin components, and destabilizes the ST state. 
As seen in Figs.~\ref{PD_GW_Omega_U}(b-c) the transition between 
the MI and ST state is discontinuous (first-order) 
for any {$\hbar\Omega/t$} when opposite spin repulsion 
$U_{\uparrow\downarrow}/U$ is large, but for small 
$U_{\uparrow\downarrow}/U$, the transition is continuous. 
In order to clarify this effect, we develop next a Ginzburg-Landau theory.

The nature of the superfluid-insulator transition 
when $\rho_\uparrow=\rho_\downarrow$ can be described by the 
Ginzburg-Landau energy 
\begin{eqnarray}
\frac{E_{\rm GL}}{M}
=
\xi({\bm k})\left(\Phi_{\rm I}^2 + \Phi_{\rm II}^2 \right) 
+\frac{\mathit{\Gamma}_1}{2}\left(\Phi_{\rm I}^4 + \Phi_{\rm II}^4\right)
+\mathit{\Gamma}_2\Phi_{\rm I}^2\Phi_{\rm II}^2
\label{ene0}
\end{eqnarray}
up to fourth order of the order parameters 
$\Phi_{\rm I} = \vert \Phi_{\bar{q}} \vert$ and 
$\Phi_{\rm II} = \vert \Phi_{-\bar{q}} \vert$, which describe the BEC with $\bm{k}=(\pm \bar{q},0,0)$. Note that the higher harmonics are negligible in the vicinity of the transition. The value of $\bar{q}$ is determined so that the function $\xi({\bm k})$ attains its minimum value $- \bar{\mu}$ at ${\bm k}=(\pm\bar{q},0,0)$. When $\bar{\mu}>0$, the bosons condense at $\bar{q}$ and/or $-\bar{q}$ with $\bar{q} \ne 0$, or simply at $\bar{q} = 0$. 
For $\mathit{\Gamma}_1<\mathit{\Gamma}_2$, 
the minimization of Eq.~({\ref{ene0}) gives $|\Phi_{\bar{q}}|\neq 0$ and 
$|\Phi_{-\bar{q}}|=0$ (or vice-versa), and 
thus the $Z_2$ symmetry related to $\bar{q}$ or $-\bar{q}$ 
is broken. In this case, the transition from 
MI to PS takes place. On the other hand, the condition $\mathit{\Gamma}_1>\mathit{\Gamma}_2$ gives $|\Phi_{\bar{q}}|=|\Phi_{-\bar{q}}|\equiv \Phi \neq 0$, resulting in the transition to the ST or CSF phase. When $\bar{q}/\pi$ is an irreducible fraction $\zeta/\eta$, the relative phase $\phi={\rm Arg}(\Phi_{\bar{q}}/\Phi_{-\bar{q}})$ is determined by the minimization of additional $\eta$-particle umklapp process, $\mathit{\Gamma}^\prime_\eta ((\Phi_{\bar{q}}^\ast)^\eta(\Phi_{-\bar{q}})^\eta+(\Phi_{-\bar{q}}^\ast)^\eta(\Phi_{\bar{q}})^\eta)\propto \cos \eta \phi$, which still has $\eta$-fold degeneracy. Thus the ST transition is associated with $U(1)\times Z_\eta$ symmetry breaking about the global and relative phases of $\Phi_{\pm\bar{q}}$.

The coefficients $\xi(q)$, $\mathit{\Gamma}_1$, 
$\mathit{\Gamma}_2$ and $\mathit{\Gamma}^\prime_\eta$ are related to the microscopic system 
parameters in the original Hamiltonian by performing a perturbative 
expansion based on a direct-product MI state. For the specific relations see supplemental 
material~\cite{supplemental-material}. We show in Fig.~\ref{PD_GW_Omega_U}(d) the values of $\mathit{\Gamma}_1$ and $\mathit{\Gamma}_2$ along the line that separates the MI phase from the others as seen in Fig.~\ref{PD_GW_Omega_U}(c). 
Note that if $\mathit{\Gamma}_1<0$ for $\mathit{\Gamma}_1<\mathit{\Gamma}_2$ or $\mathit{\Gamma}_1+\mathit{\Gamma}_2<0$ for $\mathit{\Gamma}_1>\mathit{\Gamma}_2$, the condensates have a negative compressibility, and the transition becomes first-order.

\begin{figure}[t]
\includegraphics[scale=0.45]{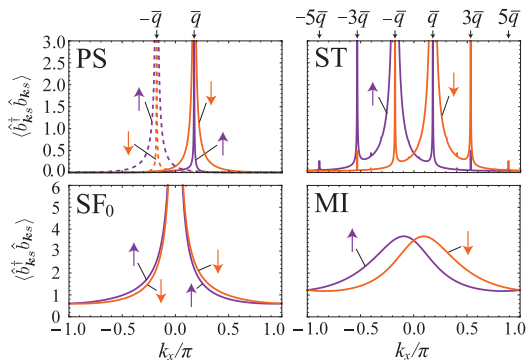}
\caption{\label{Momentum_Distribution}
(color online). The {crystal} momentum distributions $\langle \hat{b}_{\bm{k}s}^\dagger\hat{b}_{\bm{k}s}\rangle$ ($\bm{k}=(k_x,0,0)$) of the four different states along the line of {$\hbar\Omega/t=0.72$} in Fig.~\ref{PD_GW_Omega_U}(a) (at $U/t=0.5$, $1.5$, $30$, and $44$). The contribution from SF$_+$ (SF$_-$) in the PS phase is plotted by the solid (dashed) lines. } 
\end{figure}

{To assist in the experimental identification of these quantum phases, Fig.~\ref{Momentum_Distribution} shows} the {crystal} momentum distribution 
$\langle \hat{b}_{\bm{k}s}^\dagger\hat{b}_{\bm{k}s}\rangle$ at fixed density $\rho_\uparrow=\rho_\downarrow=1$. For the PS and ST states, we evaluate $\langle \hat{b}_{\bm{k}s}^\dagger\hat{b}_{\bm{k}s}\rangle$ via the Bogoliubov Hamiltonian $\hat {\mathcal H}_{B}$ at relatively weak interactions. However, for the SF$_0$ and MI states, we calculate $\langle \hat{b}_{\bm{k}s}^\dagger\hat{b}_{\bm{k}s}\rangle$ via a generalized Holstein-Primakoff approach~\cite{supplemental-material} based on the Gutzwiller variational state describing the strongly coupled regime. 
Since the PS state consists of independent domains
of SF$_+$ and SF$_-$, we plot the simple average of the two contributions. 
{The crystal momentum distribution discussed here does not include the effects of Wannier functions, but can be easily extracted from standard momentum distribution measurements.}

As seen in Figs.~\ref{Momentum_Distribution},  
the momentum distribution of the PS state exhibits two independent 
peaks around $\bm{k}=(\bar{q},0,0)$ and $\bm{k}=(-\bar{q},0,0)$, 
which come from the SF$_+$ and SF$_-$ contributions, respectively, 
while the ST state shows additional peaks due to the higher harmonics. 
The SF$_0$ state exhibits a peak around 
$\bm{k}=\bm{0}$ as in the case of a standard uniform superfluid state, 
although the reflectional symmetry with respect to $k_x\rightarrow -k_x$ 
is absent for each spin component. In the MI state, only a broad peak is 
observed at the momenta where the condensation occurs in the 
neighboring superfluid state. {The stark differences between these crystal momentum distributions also enable the direct imaging of the different phases present in inhomogeneous trapped systems.}

In summary, we {investigated} the quantum phases of two-component bosons 
in optical lattices as a function of spin-orbit coupling, Rabi frequencies 
and interactions. In phase diagrams at zero detuning, we identified four 
different regions occupied by {uniform, non-uniform and phase-separated superfluids} or Mott insulators. Finally, we characterized these phases by 
calculating their {crystal} momentum distributions, which can be easily measured 
experimentally.

{We thank N. E. Lundblad and D. Trypogeorgos for a careful reading of the manuscript. DY thanks the support of CREST, JST No. JPMJCR1673, and of KAKENHI from the Japan Society for the Promotion of Science: Grant No. 26800200. ISB thanks the support of AFOSRs Quantum Matter MURI, NIST, and the NSF through the PFC at the JQI. CARSdM acknowledges the support of JQI and NIST via its visitors program, the Galileo Galilei Institute for Theoretical Physics via a Simons Fellowship and the Aspen Center for Physics via NSF grant PHY1607611.}

\onecolumngrid

\newpage 

\subsection{Supplementary Material for ``Quantum Phases of Two-Component Bosons with Spin-Orbit Coupling in Optical Lattices''}
\renewcommand{\thesection}{\Alph{section}}
\renewcommand{\thefigure}{S\arabic{figure}}
\renewcommand{\thetable}{S\Roman{table}}
\setcounter{figure}{0}
\newcommand*{\citenamefont}[1]{#1}
\newcommand*{\bibnamefont}[1]{#1}
\newcommand*{\bibfnamefont}[1]{#1}

\def\bs{{\bf S}}
\def\bk{{\bf k}}
\def\bp{{\bf p}}
\def\bq{{\bf q}}
\def\bQ{{\bf Q}}
\def\b0{{\bf 0}}
\def\br{{\bf r}}
\def\vpa{V^{\parallel}}
\def\vpe{V^{\perp}}
\def\dag{\dagger}
\def\cM{{\cal M}}
\def\bra{\langle}
\def\ket{\rangle}
\def\bbra{\langle\!\langle}
\def\kket{\rangle\!\rangle}
\def\vev#1{\langle{#1}\rangle}
\def\emin{\epsilon_{\rm min}}
\def\non{\nonumber\\}
\renewcommand{\theequation}{S\arabic{equation}}
\renewcommand{\thesection}{\Alph{section}}
\renewcommand{\thefigure}{S\arabic{figure}}
\renewcommand{\thetable}{S\Roman{table}}
\setcounter{equation}{0}
In this Supplementary Material, we provide more technical details on the theoretical treatment of the two-component Bose-Hubbard model with spin-orbit coupling in the frameworks of (I) the Bogoliubov approach and (II) the Gutzwiller theory.
\section{\label{1}I. Bogoliubov approach} 
We use the Bogoliubov approach to provide a mean-field description of Bose-Einstein condensates (BECs), and to construct a systematic expansion in quantum fluctuations around the mean field. Under the assumption of multiple BECs with different momenta, the bosonic fields $\hat{b}_{\bm{k}s}$ can be separated in the form: 
\begin{eqnarray}
\hat{b}_{\bm{k}s}=\sum^\prime_{q}\sqrt{M}\psi_{qs}\delta_{{\bm k}=(q,0,0)}+\hat{a}_{\bm{k}s},~\label{sBogoliubov_ansatz}
\end{eqnarray}
where $\sqrt{M}\psi_{qs}$ and $\hat{a}_{\bm{k}s}$ describe condensates with momentum ${\bm k}=(q,0,0)$ and the residual bosons outside the condensates, respectively. Here, $M$ is the number of lattice sites and the sum $\sum^\prime_{q}$ runs over a set of momenta $\{ q \}$ of multiple BECs. For $U\ll t\rho$, i.e., when the particle density is very high or the interactions between particles are much weaker than the hopping amplitude, we can treat the fluctuations $\hat{a}_{\bm{k}s}$ in a perturbative fashion, and expand the Hamiltonian $\hat{\mathcal{H}}=\hat{\mathcal{H}}_0+\hat{\mathcal{H}}_{\rm int}$ into a power series of $\hat{a}_{\bm{k}s}$.

\subsection{\label{11}A. Mean-field theory}
The lowest-order terms, which involve no fluctuation operators, describe the energy of the condensates:
\begin{eqnarray}
\frac{E_0}{M}&=&\sum^\prime_{q}\left(\begin{array}{cc}
\psi_{q\uparrow}^\ast  &  \psi_{q\downarrow}^\ast\end{array}\right) \left(\begin{array}{cc}
\epsilon_{\bm{q}\uparrow}-\mu &  \Omega\\
\Omega&\epsilon_{\bm{q}\downarrow}-\mu\end{array}\right) \left(\begin{array}{c}
\psi_{q\uparrow}  \\  \psi_{q\downarrow}\end{array}\right)
\nonumber\\&&+\sum^\prime_{q_1,q_2,q_3,q_4}\delta^{(2\pi)}_{q_1+q_2,q_3+q_4}\left[\frac{U}{2}\sum_s \psi_{q_1s}^\ast\psi_{q_2s}^\ast\psi_{q_3s}\psi_{q_4s}+U_{\uparrow\downarrow} \psi_{q_1\uparrow}^\ast\psi_{q_2\downarrow}^\ast\psi_{q_3\downarrow}\psi_{q_4\uparrow}\right],\label{sGP_energy}
\end{eqnarray}
where $\bm{q}\equiv (q,0,0)$ and the Kronecker delta implements momentum conservation (modulo $2\pi$). The particle density $\rho\equiv \frac{1}{M}\sum_{{\bm k}s}\langle \hat{b}_{{\bm k}s}^\dagger\hat{b}_{{\bm k}s} \rangle$ is given by $\sum_{q}^\prime\sum_s|\psi_{qs}|^2$ at the lowest-order approximation. The minimization of $E_0$ with respect to $\psi_{qs}$ leads to the following time-independent Gross-Pitaevskii equation for the mean-field condensate configuration in the momentum space:
\begin{eqnarray}
(\epsilon_{\bm{q}s}-\mu )\psi_{qs}  +\Omega\psi_{q\bar{s}}+\sum^\prime_{q_2,q_3,q_4}\delta^{(2\pi)}_{q+q_2,q_3+q_4}\left[U\psi_{q_2s}^\ast\psi_{q_3s}\psi_{q_4s}+U_{\uparrow\downarrow} \psi_{q_2\bar{s}}^\ast\psi_{q_3\bar{s}}\psi_{q_4s}\right]=0~~({\rm for}~q\in \{q\},~s=\uparrow,\downarrow).\label{sGP_eq}
\end{eqnarray}
Here $\bar{s}=-s$.

When the set of condensate momenta $\{q\}$ consists of only a single value $\bar{q}$ corresponding, for example to phases SF$_\pm$ and SF$_0$ defined in the main text), the Gross-Pitaevskii equation is reduced to the simple form
\begin{eqnarray}
\left(\begin{array}{cc}
\epsilon_{\bar{\bm{q}}\uparrow}-\mu+U|\psi_{\bar{q}\uparrow}|^2+U_{\uparrow\downarrow}|\psi_{\bar{q}\downarrow}|^2 &  \Omega\\
\Omega&\epsilon_{\bar{\bm{q}}\uparrow}-\mu+U|\psi_{\bar{q}\downarrow}|^2+U_{\uparrow\downarrow}|\psi_{\bar{q}\downarrow}|^2\end{array}\right)\left(\begin{array}{c}
\psi_{\bar{q}\uparrow}  \\  \psi_{\bar{q}\downarrow}\end{array}\right)=\bm{0}.\label{sGP_singleq}
\end{eqnarray}
We solve the set of self-consistent equations~(\ref{sGP_singleq}) using the Newton-Raphson method under the constraint $\sum_s|\psi_{\bar{q}s}|^2=\rho$. At the same time, the condensate momentum $\bar{q}$ has to be determined such that the energy~(\ref{sGP_energy}) is also minimized with respect to $\bar{q}$.

In the striped superfluid (ST) phase, the condensates with two different momenta $-\bar{\bm{q}}_1\equiv(-\bar{q}_1,0,0)$ and $\bar{\bm{q}}_2\equiv(\bar{q}_2,0,0)$ coexist, and higher harmonics are generated at an interval of $\bar{q}_1+\bar{q}_2$ due to scattering processes induced by interactions $U$ and $U_{\uparrow\downarrow}$. Therefore, the set of condensate momenta $\{q\}$ consists of multiple components given by $-\bar{q}_1+n(\bar{q}_1+\bar{q}_2)$ where $n$ is an integer. When $(\bar{q}_1+\bar{q}_2)/2\pi$ is an irreducible fraction in the form $\zeta/\eta$, the number of independent momenta in $\{q\}$ remains finite ($=\eta$) since the momenta that lie outside the first Brillouin zone ($-\pi\leq k_x< \pi$) can be reduced to equivalent ones located inside the first zone by suitable addition or subtraction of a reciprocal lattice vector. In this case, the integer $n$ varies in the range from $n_{\rm min}$ to $n_{\rm max}$ with $(n_{\rm min},n_{\rm max})=(-\eta/2+1,\eta/2)$ for even $\eta$ and $(n_{\rm min},n_{\rm max})=(-(\eta-1)/2,(\eta-1)/2)$ for odd $\eta$. Therefore, Eq.~(\ref{sGP_eq}) becomes a set of $2\eta$ self-consistent equations, which must be solved with the density sum rule $\sum_q^\prime\sum_s|\psi_{qs}|^2=\rho$ and the minimization of the condensate energy~(\ref{sGP_energy}) with respect to the values of the fundamental momenta $\bar{q}_1$ and $\bar{q}_2$. If $\eta$ is very large or $(\bar{q}_1+\bar{q}_2)/2\pi$ is irrational (i.e. $\eta\gg 1$), we need to introduce a large number of variables $\psi_{qs}$ ($q\in \{q\}$) to solve Eq.~(\ref{sGP_eq}). In practical calculations, we truncate our system at very high harmonic components and keep a finite number of condensate mean fields to obtain well-converged energies. Note that for zero detuning $\delta=0$, the pseudospin symmetry of the system leads to  $\bar{q}_1=\bar{q}_2\equiv\bar{q}$.

\subsection{\label{12}B. Excitation spectra and momentum distribution}
The terms involving a single fluctuation operator $\hat{a}_{\bm{k}s}$ (or $\hat{a}_{\bm{k}s}^\dagger$) vanish when the solution of Eq.~(\ref{sGP_eq}) is substituted into $\psi_{qs}$. Therefore, the first correction to the mean-field theory arises from the quadratic terms
\begin{eqnarray}
&&\sum_{\bm{k}}\left(\begin{array}{cc}
\hat{a}_{\bm{k}\uparrow}^\dagger  &  \hat{a}_{\bm{k}\downarrow}^\dagger\end{array}\right) \left(\begin{array}{cc}
\epsilon_{\bm{k}\uparrow}-\mu &  \Omega\\
\Omega&\epsilon_{\bm{k}\downarrow}-\mu\end{array}\right) \left(\begin{array}{c}
\hat{a}_{\bm{k}\uparrow}  \\  \hat{a}_{\bm{k}\downarrow}\end{array}\right)+\frac{1}{2}\sum_{q_1,q_2}^\prime\sum_{{\bm k}_1,{\bm k}_2}\Bigg[\nonumber\\
&&\delta^{(2\pi)}_{k_{1x}+q_1,k_{2x}+q_2}\delta_{k_{1y},k_{2y}}\delta_{k_{1z},k_{2z}}\sum_s\Big(4U\psi_{q_1s}^\ast\psi_{q_2s}\hat{a}_{\bm{k}_1s}^\dagger \hat{a}_{\bm{k}_2s}+2U_{\uparrow\downarrow} (\psi_{q_1s}^\ast\psi_{q_2\bar{s}}\hat{a}_{\bm{k}_1\bar{s}}^\dagger \hat{a}_{\bm{k}_2s}+\psi_{q_1s}^\ast\psi_{q_2s}\hat{a}_{\bm{k}_1\bar{s}}^\dagger \hat{a}_{\bm{k}_2\bar{s}})\Big)\nonumber\\
&&+\delta^{(2\pi)}_{k_{1x}-q_1,q_2-k_{2x}}\delta_{k_{1y},-k_{2y}}\delta_{k_{1z},-k_{2z}}\Big(U\sum_s\psi_{q_1s}^\ast\psi_{q_2s}^\ast\hat{a}_{\bm{k}_1s} \hat{a}_{\bm{k}_2s}+2U_{\uparrow\downarrow} \psi_{q_1\uparrow}^\ast\psi_{q_2\downarrow}^\ast\hat{a}_{\bm{k}_1\downarrow} \hat{a}_{\bm{k}_2\uparrow}+{\rm H.c.}\Big)\Bigg],
\end{eqnarray}
which can be rewritten in a simpler form:
\begin{eqnarray}
\frac{1}{2}\sum_{\bm{k}}(\hat{\bm{a}}^\dagger_{\bm{k}}~(\hat{\bm{a}}_{-\bm{k}})^T) \left(\begin{array}{cc}
\mathbf{A}_{\bm{k}} & \mathbf{B}_{\bm{k}}\\
\mathbf{B}^\ast_{-\bm{k}} &\mathbf{A}_{-\bm{k}}^\ast \end{array}\right) 
\left(\begin{array}{c}
\hat{\bm{a}}_{\bm{k}}\\
(\hat{\bm{a}}_{-\bm{k}}^\dagger)^T\end{array}\right)-\frac{1}{2}\sum_{\bm{k}}{\rm Tr}\left(\mathbf{A}_{-\bm{k}}^\ast\right).\label{sBogoliubov_Hamiltonian}
\end{eqnarray}

For the SF$_\pm$ or SF$_0$ phases with a single condensate momentum $\bar{\bm{q}}=(\bar{q},0,0)$, the column vector $\hat{\bm{a}}_{\bm{k}}$ and the matrices $\mathbf{A}_{\bm{k}},\mathbf{B}_{\bm{k}}$ are given by $\hat{\bm{a}}_{\bm{k}}=(\hat{a}_{\bar{\bm{q}}+\bm{k}\uparrow},\hat{a}_{\bar{\bm{q}}+\bm{k}\downarrow})^T$ and
\begin{eqnarray*}
\mathbf{A}_{\bm{k}}=\left(\begin{array}{cc}
\epsilon_{\bar{\bm{q}}+\bm{k}\uparrow}-\mu+2U|\psi_{\bar{q}\uparrow}|^2+U_{\uparrow\downarrow}|\psi_{\bar{q}\downarrow}|^2 &  \Omega+ U_{\uparrow\downarrow }\psi_{\bar{q}\downarrow}^\ast\psi_{\bar{q}\uparrow}\\
\Omega+ U_{\uparrow\downarrow} \psi_{\bar{q}\uparrow}^\ast\psi_{\bar{q}\downarrow}&\epsilon_{\bar{\bm{q}}+\bm{k}\downarrow}-\mu+2U|\psi_{\bar{q}\downarrow}|^2+U_{\uparrow\downarrow}|\psi_{\bar{q}\uparrow}|^2\end{array}\right),~
\mathbf{B}_{\bm{k}}=\left(\begin{array}{cc}
U \psi_{\bar{q}\uparrow}^2 & U_{\uparrow\downarrow} \psi_{\bar{q}\downarrow}\psi_{\bar{q}\uparrow} \\
U_{\uparrow\downarrow} \psi_{\bar{q}\uparrow}\psi_{\bar{q}\downarrow}& U \psi_{\bar{q}\downarrow}^2 \end{array}\right).
\end{eqnarray*}
For the striped superfluid (ST) phase in which $(\bar{q}_1+\bar{q}_2)/2\pi$ is a rational number $\zeta/\eta$, the column vector $\hat{\bm{a}}_{\bm{k}}$ consists of $2\eta$ components: 
\begin{eqnarray*}
\hat{\bm{a}}_{\bm{k}}=(\hat{a}_{-\bar{\bm{q}}_1+n_{\rm min}(\bar{\bm{q}}_1+\bar{\bm{q}}_2)+\bm{k}\uparrow},~\cdots,\hat{a}_{-\bar{\bm{q}}_1+n(\bar{\bm{q}}_1+\bar{\bm{q}}_2)+\bm{k}\uparrow},\hat{a}_{-\bar{\bm{q}}_1+n(\bar{\bm{q}}_1+\bar{\bm{q}}_2)+\bm{k}\downarrow}  ,~\cdots,\hat{a}_{-\bar{\bm{q}}_1+n_{\rm max}(\bar{\bm{q}}_1+\bar{\bm{q}}_2)+\bm{k}\downarrow})^T
\end{eqnarray*}
with integers $n\in[n_{\rm min},n_{\rm max}]$. The $2\eta\times 2\eta$ matrices $\mathbf{A}_{\bm{k}}$ and $\mathbf{B}_{\bm{k}}$ are given by
\begin{eqnarray*}
\mathbf{A}_{\bm{k}}=\left(\begin{array}{ccccc}
\bar{\mathbf{A}}_{n_{\rm min},n_{\rm min}}&\cdots & \bar{\mathbf{A}}_{n_{\rm min},n}&\cdots &\bar{\mathbf{A}}_{n_{\rm min},n_{\rm max}}\\
\vdots &\ddots& \vdots &\iddots &\vdots \\
\bar{\mathbf{A}}_{n,n_{\rm min}} &\cdots & \bar{\mathbf{A}}_{n,n}& \cdots&\bar{\mathbf{A}}_{n,n_{\rm max}}\\
\vdots &\iddots& \vdots &\ddots &\vdots \\
\bar{\mathbf{A}}_{n_{\rm max},n_{\rm min}}&\cdots & \bar{\mathbf{A}}_{n_{\rm max},n}&\cdots &\bar{\mathbf{A}}_{n_{\rm max},n_{\rm max}}\end{array}\right)~{\rm and}~\mathbf{B}_{\bm{k}}=\left(\begin{array}{ccccc}
\bar{\mathbf{B}}_{n_{\rm min}n_{\rm min}}&\cdots & \bar{\mathbf{B}}_{n_{\rm min},n}&\cdots &\bar{\mathbf{B}}_{n_{\rm min},n_{\rm max}}\\
\vdots &\ddots& \vdots &\iddots &\vdots \\
\bar{\mathbf{B}}_{n,n_{\rm min}} &\cdots & \bar{\mathbf{B}}_{n,n}& \cdots&\bar{\mathbf{B}}_{n,n_{\rm max}}\\
\vdots &\iddots& \vdots &\ddots &\vdots \\
\bar{\mathbf{B}}_{n_{\rm max},n_{\rm min}}&\cdots & \bar{\mathbf{B}}_{n_{\rm max},n}&\cdots &\bar{\mathbf{B}}_{n_{\rm max},n_{\rm max}}\end{array}\right)
\end{eqnarray*}
with
\begin{eqnarray*}
\bar{\mathbf{A}}_{n,n^\prime}&\equiv&\left(\begin{array}{cc}
\epsilon_{-\bar{\bm{q}}_1+n(\bar{\bm{q}}_1+\bar{\bm{q}}_2)+\bm{k}\uparrow}-\mu &  \Omega\\
\Omega&\epsilon_{-\bar{\bm{q}}_1+n(\bar{\bm{q}}_1+\bar{\bm{q}}_2)+\bm{k}\downarrow}-\mu\end{array}\right)\delta_{n,n^\prime}\\
&&+\sum_{q,q^\prime}^\prime \delta^{(2\pi)}_{q-q^\prime,-(n-n^\prime)(\bar{q}_1+\bar{q}_2)} \left(\begin{array}{cc}
2U \psi_{q\uparrow}^\ast\psi_{q^\prime\uparrow}+U_{\uparrow\downarrow} \psi_{q\downarrow}^\ast\psi_{q^\prime\downarrow}& U_{\uparrow\downarrow} \psi_{q\downarrow}^\ast\psi_{q^\prime\uparrow} \\
U_{\uparrow\downarrow} \psi_{q\uparrow}^\ast\psi_{q^\prime\downarrow}& 2U \psi_{q\downarrow}^\ast\psi_{q^\prime\downarrow} +U_{\uparrow\downarrow} \psi_{q\uparrow}^\ast\psi_{q^\prime\uparrow}\end{array}\right)~{\rm and}\\
\bar{\mathbf{B}}_{n,n^\prime}&\equiv& \sum_{q,q^\prime}^\prime \delta^{(2\pi)}_{q+q^\prime+2\bar{q}_1,(n+n^\prime)(\bar{q}_1+\bar{q}_2)} \left(\begin{array}{cc}
U \psi_{q\uparrow}\psi_{q^\prime\uparrow}& U_{\uparrow\downarrow} \psi_{q\downarrow}\psi_{q^\prime\uparrow} \\
U_{\uparrow\downarrow} \psi_{q\uparrow}\psi_{q^\prime\downarrow}& U \psi_{q\downarrow}\psi_{q^\prime\downarrow} \end{array}\right).
\end{eqnarray*}
Note again that when $\eta$ is very large or $(\bar{q}_1+\bar{q}_2)/2\pi$ is irrational, we truncate the higher harmonic components that give no meaningful contribution to the result.

The operator part of Eq.~(\ref{sBogoliubov_Hamiltonian}) can be numerically diagonalized by the generalized Bogoliubov transformation $(\hat{\bm{a}}^\dagger_{\bm{k}}~(\hat{\bm{a}}_{-\bm{k}})^T)=(\hat{\bm{\alpha}}^\dagger_{\bm{k}}~(\hat{\bm{\alpha}}_{-\bm{k}})^T)\mathbf{P}_{\bm  k}^\dagger$ with a paraunitary matrix $\mathbf{P}_{\bm  k}$~\cite{colpa-78} as 
\begin{eqnarray}
\frac{1}{2}\sum_{\bm{k}}(\hat{\bm{\alpha}}^\dagger_{\bm{k}}~(\hat{\bm{\alpha}}_{-\bm{k}})^T) \mathbf{P}_{\bm  k} \left(\begin{array}{cc}
\mathbf{A}_{\bm{k}} & \mathbf{B}\\
\mathbf{B}^\ast &\mathbf{A}_{-\bm{k}}^\ast \end{array}\right)\mathbf{P}_{\bm  k}^\dagger
\left(\begin{array}{c}
\hat{\bm{\alpha}}_{\bm{k}}\\
(\hat{\bm{\alpha}}_{-\bm{k}}^\dagger)^T\end{array}\right)&=&
\frac{1}{2}\sum_{\bm{k}}(\hat{\bm{\alpha}}^\dagger_{\bm{k}}~(\hat{\bm{\alpha}}_{-\bm{k}})^T) \left(\begin{array}{cc}
\mbox{\boldmath $\omega$}_{\bm{k}} &\mbox{\boldmath $0$} \\
\mbox{\boldmath $0$}&\mbox{\boldmath $\omega$}_{-\bm{k}} \end{array}\right)
\left(\begin{array}{c}
\hat{\bm{\alpha}}_{\bm{k}}\\
(\hat{\bm{\alpha}}_{-\bm{k}}^\dagger)^T\end{array}\right)\nonumber\\
&=&\sum_{\bm k}\hat{\bm{\alpha}}^\dagger_{\bm{k}} \mbox{\boldmath $\omega$}_{\bm{k}}\hat{\bm{\alpha}}_{\bm{k}}+\frac{1}{2}\sum_{\bm k}{\rm Tr}\left(\mbox{\boldmath $\omega$}_{-\bm{k}}\right),\label{sgBT_1}
\end{eqnarray}
where $\mbox{\boldmath $\omega$}_{\bm{k}}$ is a diagonal matrix and $\mbox{\boldmath $0$}$ is the null matrix. The paraunitary matrix $\mathbf{P}_{\bm  k}$ satisfies the relations
\begin{eqnarray}
\left(\begin{array}{cc}
\mathbf{A}_{\bm{k}} & \mathbf{B}_{\bm{k}}\\
\mathbf{B}^\ast_{-\bm{k}} &\mathbf{A}_{-\bm{k}}^\ast \end{array}\right) \mathbf{P}_{\bm  k}={\bf \Sigma}\mathbf{P}_{\bm  k}\left(\begin{array}{cc}
\mbox{\boldmath $\omega$}_{\bm{k}} & \mbox{\boldmath $0$}\\
\mbox{\boldmath $0$}&-\mbox{\boldmath $\omega$}_{-\bm{k}} \end{array}\right),~\mathbf{P}_{\bm  k}^\dagger {\bf \Sigma} \mathbf{P}_{\bm  k}={\bf \Sigma},~{\rm and}~\mathbf{P}_{\bm  k} {\bf \Sigma} \mathbf{P}_{\bm  k}^\dagger={\bf \Sigma},~\label{sgBT_2}
\end{eqnarray}
where $\bf{\Sigma}$ is the diagonal matrix of dimensions $2\eta \times 2\eta$, whose diagonal elements are $1$ for the upper-half and $-1$ for the lower-half entries. The latter two conditions are required such that the transformed operators $\hat{\bm{\alpha}}_{\bm{k}}$ still obey bosonic commutation relations.

The elements of the diagonal matrix $\mbox{\boldmath $\omega$}_{\bm{k}}$ give the excitation spectra of the Bogoliubov quasiparticles. Moreover, using the transformation coefficients $\mathbf{P}_{\bm  k}$ and the relation $\langle \hat{\bm{\alpha}}_{\bm{k}} \hat{\bm{\alpha}}_{\bm{k}}^\dagger\rangle=\mbox{\boldmath $1$}$ (at zero temperature) with $\mbox{\boldmath $1$}$ denoting the identity matrix, one can evaluate the fluctuation part $\langle \hat{a}_{\bm{k}s}^\dagger\hat{a}_{\bm{k}s}\rangle$ of the momentum distribution
\begin{eqnarray}
\langle \hat{b}_{\bm{k}s}^\dagger\hat{b}_{\bm{k}s}\rangle=M\sum^\prime_{q}|\psi_{qs}|^2\delta_{{\bm k}={\bm q}}+\langle \hat{a}_{\bm{k}s}^\dagger\hat{a}_{\bm{k}s}\rangle.
\end{eqnarray}
Now that we have discussed the technical details of the regime where the particle density is high or the interaction energy is much smaller than the hopping, we are ready to discuss next the opposite regime, where the Gutzwiller theory is a more suitable approach.

\section{\label{2}II. Gutzwiller theory}
The Hamiltonian $\hat{\mathcal{H}}=\hat{\mathcal{H}}_0+\hat{\mathcal{H}}_{\rm int}$ considered in the main text is written in real space as
\begin{eqnarray}
\hat{\mathcal{H}}_0&=&-\sum_{ijs}t_{ij}^s \hat{b}_{is}^\dagger\hat{b}_{js}+\Omega \sum_{i}\left(\hat{b}_{i\uparrow}^\dagger\hat{b}_{i\downarrow}+\hat{b}_{i\downarrow}^\dagger\hat{b}_{i\uparrow}\right)+\delta\sum_i (\hat{n}_{i\uparrow}-\hat{n}_{i\downarrow})-\mu\sum_{is}\hat{n}_{is},\label{sH0}\\
\hat{\mathcal{H}}_{\rm int}&=&\frac{U}{2}\sum_{is}\hat{n}_{is}(\hat{n}_{is}-1)+U_{\uparrow\downarrow}\sum_i \hat{n}_{i\uparrow}\hat{n}_{i\downarrow},\label{sHint}
\end{eqnarray}
where $\hat{n}_{is} = \hat{b}_{is}^\dagger\hat{b}_{is}$. The hopping term depends on the spin state $s=+$ ($\uparrow$) or $-$ ($\downarrow$) as
\begin{eqnarray}
t_{ij}^s=\left\{ \begin{array}{ll}
te^{is\bm{k}_T\cdot(\bm{r}_j-\bm{r}_i)}  &  (|\bm{r}_j-\bm{r}_i)|=1) \\
0 & ({\rm otherwise})
\end{array} \right. ,
\end{eqnarray}
where $\bm{k}_T=({k_T},0,0)$ is the momentum transfer and 
$\bm{r}_i=(x_i,y_i,z_i)$ is the three dimensional coordinate at lattice 
site $i$, with lattice constant set to be $1$.

In order to describe Mott insulator (MI) transitions at strong interactions, we need to keep all onsite terms of the Hamiltonian defined in Eqs. (\ref{sH0}) and (\ref{sHint}), but we can decouple the hopping term as 
\begin{eqnarray}
-\sum_{ijs}t_{ij}^s \hat{b}_{is}^\dagger\hat{b}_{js}\approx  -\sum_{ijs}\left(t_{ji}^s \psi_{js}^{\ast} \hat{b}_{is}+t_{ij}^s \psi_{js} \hat{b}_{is}^\dagger-t_{ij}^s \psi_{is}^{\ast}\psi_{js}\right),\label{sGW_decoupling}
\end{eqnarray}
where $\psi_{is}\equiv\langle \hat{b}_{is} \rangle$ are Gutzwiller-type variational fields. This approximation is equivalent to approximating the ground-state wavefunction by a direct product state in real space. The Gutzwiller fields play the role of order parameter for superfluid states. The total Hamiltonian of the system becomes a sum of effective local Hamiltonians: $\hat{\mathcal{H}}\approx\sum_i \left(\hat{{h}}^{\rm GW}_i+\sum_{js}t_{ij}^s \psi_{is}^{\ast}\psi_{js}\right)$ with 
\begin{eqnarray}
\hat{{h}}^{\rm GW}_i&=& -\sum_{js}\left(t_{ji}^s \psi_{js}^{\ast} \hat{b}_{is}+t_{ij}^s \psi_{js} \hat{b}_{is}^\dagger\right)-\mu\left(\hat{n}_{i\uparrow}+\hat{n}_{i\downarrow}\right)+\Omega(\hat{b}^\dagger_{i\uparrow}\hat{b}_{i\downarrow}+\hat{b}_{i\downarrow}^\dagger\hat{b}_{i\uparrow})\nonumber\\&&
+\frac{U}{2}\left(\hat{n}_{i\uparrow}(\hat{n}_{i\uparrow}-1)+\hat{n}_{i\downarrow}(\hat{n}_{i\downarrow}-1)\right)+U_{\uparrow\downarrow}\hat{n}_{i\uparrow}\hat{n}_{i\downarrow}\label{shlocal}.
\end{eqnarray}
Next, we analyze in detail the Gutzwiller ground state.

\subsection{\label{21}A. Gutzwiller ground state}
The effective local Hamiltonian $\hat{{h}}^{\rm GW}_i$ is now given only in terms of local operators $\hat{b}_{is}$ and $\hat{n}_{is}$ at site $i$. Nevertheless, the local Hamiltonians are still coupled with each other since the coefficients contain the Gutzwiller variational fields $\psi_{js}$ of the neighboring sites. The value of $\psi_{is}$ at each site $i$ must be determined self-consistently from the condition  $\psi_{is}=\langle \hat{b}_{is} \rangle$ in the ground state of the local Hamiltonian $\hat{{h}}^{\rm GW}_i$.

For the SF$_\pm$ or SF$_0$ phases, the mean fields have the form
\begin{eqnarray}
\psi_{is}=\psi_{\bar{q}s}e^{i\bar{q}x_i}
\end{eqnarray}
with a single condensate momentum $\bar{\bm{q}}=(\bar{q},0,0)$. Therefore, under the gauge transformation $\tilde{b}_{is}\equiv \hat{b}_{is}e^{-i\bar{q}x_i}$ the hopping part of the effective Hamiltonian can be rewritten as
\begin{eqnarray}
-2t\left(\cos(sk_T+\bar{q}\right)+2)\left(\psi_{\bar{q}s}^{\ast} \tilde{b}_{is}+\psi_{\bar{q}s} \tilde{b}_{is}^\dagger\right).\label{stilde_ht}
\end{eqnarray}
The remaining part of $\hat{{h}}^{\rm GW}_i$ keeps its original form since the phase factor $e^{i\bar{q}x_i}$ is cancelled out. In the tilde representation, the coefficients of the operators become all site-independent, and thus the effective local Hamiltonians at different sites are now completely decoupled and equivalent.

In order to obtain the values of the Gutzwiller fields $\psi_{\bar{q}s}$, we need to carry out a self-consistent calculation at a single site $i$. For this purpose, we use the following procedure: (i) start with certain initial values for $\psi_{\bar{q}\uparrow}$ and $\psi_{\bar{q}\downarrow}$; (ii) express $\hat{{h}}^{\rm GW}_i$ in the tilde representation as a matrix $\mathbf{{h}}^{\rm GW}_i$ in the two-component Fock basis $|n_\uparrow,n_\downarrow\rangle=|0,0\rangle,|1,0\rangle,|0,1\rangle,|2,0\rangle,|1,1\rangle,|0,2\rangle,\cdots$, where $n_s$ is the occupation number of each pseudospin; (iii) numerically diagonalize $\mathbf{{h}}^{\rm GW}_i$ and evaluate the expectation values $\langle \tilde{b}_{is} \rangle$ in the lowest eigenstate; (iv) update $\psi_{\bar{q}s}$ with $\langle \tilde{b}_{is} \rangle$; (v) repeat (ii-iv) until self-consistency is reached.

In (ii) and (iii), we make a restriction on $n_\uparrow$ and $n_\downarrow$ to a maximum occupation of $12$ throughout the present work in order to truncate the infinite Hilbert space of bosons. The matrix $\mathbf{{h}}^{\rm GW}_i$ is diagonalized with a unitary matrix $\mathbf{R}$ giving
\begin{eqnarray}
\mathbf{R}^\dagger \mathbf{{h}}^{\rm GW}_i \mathbf{R}=\left(\begin{array}{cccc}
\epsilon_0 & &&\\
&\epsilon_1 &&\\
&& \ddots &\\
&&&\epsilon_{n_{\rm tr}}\end{array}\right) \label{sRmat}
\end{eqnarray}
where we choose $\mathbf{R}$ such that the eigenvalues $\epsilon_0,\epsilon_1,\cdots \epsilon_{n_{\rm tr}}$ are sorted in ascending order. The number of excited states $n_{\rm tr}$ depends on the truncation of the local Hilbert space. The expectation values $\langle \tilde{b}_{is} \rangle$ are given by $[\mathbf{R}^\dagger \mathbf{\tilde{b}}_{is} \mathbf{R}]_{11}$ with $\mathbf{\tilde{b}}_{is}$ being the matrix representation of $\tilde{b}_{is}$. We repeat the diagonalization procedure (ii-iv) until the difference between the input and output values of $\psi_{\bar{q}s}$ becomes negligible. To accelerate the convergence, we employ the Newton-Raphson technique. In parallel with the self-consistent calculation, we also determine the value of $\bar{q}$ from the minimization of the ground-state energy per site, which is the sum of the lowest eigenvalue $\epsilon_0$ and the c-number term $\sum_{js}t_{ij}^s \psi_{is}^{\ast}\psi_{js}= 2t\sum_s\left(\cos (sk_T+\bar{q})+2\right)|\psi_{\bar{q}s}|^2$ originated from the decoupling (\ref{sGW_decoupling}).

On the other hand, for the striped superfluid (ST) phase, we cannot transform the effective Hamiltonian $\hat{{h}}^{\rm GW}_i$ into a form where the coefficients are site-independent. Hence, we have to solve simultaneously a set of single-site problems coupled through the mean fields $\psi_{is}$. The Gutzwiller fields for the ST phase are simply $\psi_{is}=\sum_q^\prime \psi_{qs}e^{iqx_i}$. The sum $\sum_q^\prime$ runs over a set of condensate momenta $\{q\}$, whose components are given by $-\bar{q}_1+n(\bar{q}_1+\bar{q}_2)$, where $n$ is an integer. When $(\bar{q}_1+\bar{q}_2)/2\pi$ is an irreducible fraction $\zeta/\eta$, the number of independent mean fields is $2\eta$ with the factor of $2$ coming from the spin index. Notice that the means fields are periodic along the x direction, that is $\psi_{(x_i+\eta,y_i,z_i)s}=\psi_{(x_i,y_i,z_i)s}$, while they are uniform  along the $y$ and $z$ directions. Therefore, we diagonalize simultaneously the $\eta$ different local Hamiltonians, which are functions of the mean fields $\psi_{is}=\sum_q^\prime \psi_{qs}e^{iqx_i}$, and calculate the expectation values $\langle \hat{b}_{is}\rangle$ at each site. The cycle is repeated until the self-consistent condition $\psi_{qs}=\frac{1}{\eta}\sum_{x_i=1}^{\eta}\langle \hat{b}_{is}\rangle e^{-iq x_i}$ ($q\in \{q\}$) is achieved. The values of the fundamental momenta $\bar{q}_1$ and $\bar{q}_2$ have to be determined by minimizing the ground-state energy. In the present work, we consider the ST state with periodicity up to $\eta=2\times 10^3$, and thus the interval of possible momenta is given by $\delta k_x=2\pi/\eta\sim 0.001\pi$, corresponding to the momentum resolution between two consecutive momenta along the $x$ direction.

\subsection{\label{22}B. Excitation spectra and momentum distribution} 
In the following, we present a treatment of non-local correlation effects beyond the Gutzwiller approximation. For simplicity, we restrict ourselves to the specific procedure only for single-$q$ states, SF$_\pm$ ($\bar{q}\neq 0$) and SF$_0$ ($\bar{q}=0$), although our method can be generalized in a straightforward way for multi-$q$ states. Using the unitary matrix $\mathbf{R}$ [Eq. (\ref{sRmat})] that diagonalizes the Gutzwiller Hamiltonian matrix $\mathbf{{h}}^{\rm GW}_i$, we introduce a multi-flavor Schwinger-boson representation of local operators $\hat{\mathcal{O}}_i=\tilde{b}_{is}$, $\tilde{b}_{is}^\dagger\tilde{b}_{is^\prime}$, $\tilde{n}_{is}$, and $\tilde{n}_{is}\tilde{n}_{is^\prime}$: 
\begin{eqnarray}
\hat{\mathcal{O}}_i=(\hat{a}_{i,0}^\dagger ~  \hat{\bm{a}}_i^\dagger)\mathbf{R}^\dagger \mathbf{O}_i \mathbf{R}\left(\begin{array}{c} \hat{a}_{i,0}\\   \hat{\bm{a}}_i \end{array}\right),\label{sSB_representation}
\end{eqnarray}
where $\hat{a}_{i,0}$ and $\hat{\bm{a}}_i=(\hat{a}_{i,1},\hat{a}_{i,2},\cdots,\hat{a}_{i,n_{\rm tr}})^T$ are Schwinger bosons and $\mathbf{O}_i$ is the matrix representation of the operator $\hat{\mathcal{O}}_i$ in the two-component Fock basis. The operator $\hat{a}_{i,0}^\dagger$ describes the creation of the local Gutzwiller ground state (the lowest eigenstate of $\mathbf{{h}}^{\rm GW}_i$), while the other Schwinger bosons $\hat{a}_{i,n\neq 0}^\dagger$ create the other (higher) eigenstates and thus describe the fluctuations around the Gutzwiller ground state. The physical subspace of states is obtained by imposing the constraint
\begin{eqnarray}
\hat{a}_{i,0}^\dagger \hat{a}_{i,0}+ \hat{\bm{a}}_i^\dagger \hat{\bm{a}}_i =1.
\label{sconstraint}
\end{eqnarray}

We rewrite the original Hamiltonian $\mathcal{H}=\mathcal{H}_0+\mathcal{H}_{\rm int}$ in terms of Schwinger bosons by substituting Eq.~(\ref{sSB_representation}) into Eqs.~(\ref{sH0}) and~(\ref{sHint}). If we take $\hat{a}_{i,0}^\dagger \hat{a}_{i,0}=1$ and $\hat{a}_{i,n}^\dagger \hat{a}_{i,n}=0$ for $n\neq 0$, we reproduce the Gutzwiller ground-state energy. In order to take into account fluctuations around the Gutzwiller ground state, we expand the Hamiltonian into a power series about the fluctuation operators $ \hat{a}_{i,n\neq 0}$. To that end, we eliminate the operator $\hat{a}_{i,0}$ via the constraint (\ref{sconstraint}) leading to the relations 
\begin{eqnarray}
\hat{a}_{i,0}^\dagger \hat{a}_{i,0}\rightarrow 1-\hat{\bm{a}}_i^\dagger \hat{\bm{a}}_i~~{\rm and}~~\hat{a}_{i,0}^\dagger \hat{a}_{i,n}\rightarrow \sqrt{1-\hat{\bm{a}}_i^\dagger \hat{\bm{a}}_i}\hat{a}_{i,n}=\left(1-\frac{1}{2}\hat{\bm{a}}_i^\dagger \hat{\bm{a}}_i+\cdots\right)\hat{a}_{i,n}~{\rm for}~n\neq 0.\label{sSB_expansion}
\end{eqnarray}
The expansion of the square root can be justified when the fluctuation $\hat{\bm{a}}_i^\dagger \hat{\bm{a}}_i$ is sufficiently  small.

After the expansion, the terms that are linear in $\hat{a}_{i,n\neq 0}$ and $\hat{a}_{i,n\neq 0}^\dagger$ vanish since the ground-state energy at the level of the Gutzwiller approximation (the zeroth order terms of the expansion) is already minimized by substituting the self-consistently converged values of $\psi_{\bar{q}s}$. Thus the first correction to the Gutzwiller approximation arises from the quadratic terms. The quadratic Hamiltonian can be written in the same form as in Bogoliubov theory [Eq.~(\ref{sBogoliubov_Hamiltonian})] after the Fourier transformation, where the corresponding matrices $\mathbf{A}_{\bm k}$ and $\mathbf{B}_{\bm k}$ in this case are derived from Eqs.~(\ref{sSB_representation}) and (\ref{sSB_expansion}). The excitation spectrum is obtained by the generalized Bogoliubov transformation $(\hat{\bm{a}}^\dagger_{\bm{k}}~(\hat{\bm{a}}_{-\bm{k}})^T)=(\hat{\bm{\alpha}}^\dagger_{\bm{k}}~(\hat{\bm{\alpha}}_{-\bm{k}})^T)\mathbf{P}_{\bm  k}^\dagger$ [Eqs.~(\ref{sgBT_1}) and (\ref{sgBT_2})]. The momentum distribution of the particles
\begin{eqnarray}
\langle \hat{b}_{\bm{k}s}^\dagger\hat{b}_{\bm{k}s}\rangle=\frac{1}{M}\sum_{ij}\langle \tilde{b}_{is}^\dagger\tilde{b}_{js}\rangle e^{i(\bm{k}-\bar{\bm{q}})\cdot(\bm{r}_i-\bm{r}_j)}
\end{eqnarray}
is also calculated in the following way: (i) express the operator $\langle\tilde{b}_{is}^\dagger\tilde{b}_{js}\rangle$ in the Schwinger-boson representation (\ref{sSB_representation}); (ii) eliminate $\hat{a}_{i,0}$ using Eq.~(\ref{sSB_expansion}) and keep only the terms up to second order in the fluctuation operator $\hat{a}_{i,n\neq 0}$; (iii) evaluate the expectation values such as $\langle \hat{a}_{{\bm k},n}^\dagger\hat{a}_{{\bm k},n^\prime}\rangle$ using the transformation coefficients $\mathbf{P}_{\bm  k}$ and the relation $\langle \hat{\bm{\alpha}}_{\bm{k}} \hat{\bm{\alpha}}_{\bm{k}}^\dagger\rangle=\mathbf{1}$.

\subsection{\label{23}C. Ginzburg-Landau theory} 
Finally, we present the details of the Ginzburg-Landau description of the Mott-Insulator (MI) transition when the spin population is the same, that is, $\rho_\uparrow=\rho_\downarrow$ ($\delta=0$). The effective Hamiltonian within the Gutzwiller approximation is $\hat{\mathcal{H}}\approx \sum_i\left(\hat{{h}}_i^{\rm GW}+\sum_{js}t_{ij}^s \psi_{is}^{\ast}\psi_{js}\right)$ and can be separated as $\sum_i \hat{{h}}_i^{\rm GW}=\sum_i \hat{{h}}_i^{(0)}+\sum_i\hat{{h}}^t_i$ with the local fields containing the interactions being
\begin{eqnarray}
\hat{{h}}_i^{(0)}=-\mu\left(\hat{n}_{i\uparrow}+\hat{n}_{i\downarrow}\right)+\Omega(\hat{b}^\dagger_{i\uparrow}\hat{b}_{i\downarrow}+\hat{b}_{i\downarrow}^\dagger\hat{b}_{i\uparrow})+\frac{U}{2}\left(\hat{n}_{i\uparrow}(\hat{n}_{i\uparrow}-1)+\hat{n}_{i\downarrow}(\hat{n}_{i\downarrow}-1)\right)+U_{\uparrow\downarrow}\hat{n}_{i\uparrow}\hat{n}_{i\downarrow}\label{shlocal}
\end{eqnarray}
and with the local fields containing the kinetic terms being
\begin{eqnarray}
\hat{{h}}^t_i= -\sum_{js}\left(t_{ji}^s \psi_{js}^{\ast} \hat{b}_{is}+t_{ij}^s \psi_{js} \hat{b}_{is}^\dagger\right)\label{sht}.
\end{eqnarray} 
In the vicinity of the MI transition, where the Gutzwiller fields $\psi_{is}$ are small, we can treat $\sum_i \hat{{h}}^t_i$ as a perturbation about the Hamiltonian $\sum_i \hat{{h}}^{(0)}_i$. To solve the eigenvalue problem for $\hat{{h}}^{(0)}_i$
\begin{eqnarray}
\hat{{h}}^{(0)}_i|\varphi_\rho^\lambda \rangle=\varepsilon_{\rho}^{\lambda}|\varphi_\rho^\lambda \rangle,
\end{eqnarray}
we diagonalize the matrix form of $\hat{{h}}^{(0)}_i$ in each sector of the Hilbert space with filling factor $\rho$. The $\lambda$-th eigenstate $|\varphi_\rho^\lambda \rangle$ for filling $\rho$ ($\lambda=0,1,\cdots,\rho$) is given by a superposition of the local Fock states $|n_\uparrow,n_\downarrow\rangle$ that satify $n_\uparrow+n_\downarrow=\rho$:
\begin{eqnarray}
|\varphi_\rho^\lambda \rangle=\sum_{n_\uparrow+n_\downarrow=\rho}u_{(n_\uparrow,n_\downarrow)}^\lambda |n_\uparrow,n_\downarrow\rangle.
\end{eqnarray}
When $\rho=2$, for example, the eigenenergies $\varepsilon_{\rho=2}^{\lambda}$ and the corresponding eigenvectors $\bm{u}^\lambda_{\rho=2}=(u_{(2,0)}^\lambda,u_{(1,1)}^\lambda,u_{(0,2)}^\lambda)^{T}$ are
\begin{eqnarray}
\begin{array}{lllllll}
\varepsilon_{2}^{0}&=&\frac{U+U_{\uparrow\downarrow}}{2}-\sqrt{\left(\frac{U-U_{\uparrow\downarrow}}{2}\right)^2+4\Omega^2}-2\mu;&&\bm{u}^0_2&=&\frac{(-2\sqrt{2}\Omega,U-U_{\uparrow\downarrow}+\varepsilon_{2}^{2}-\varepsilon_{2}^{0},-2\sqrt{2}\Omega)^{T}}{\sqrt{16\Omega^2+(U-U_{\uparrow\downarrow}+\varepsilon_{2}^{2}-\varepsilon_{2}^{0})^2}},\\
\varepsilon_{2}^{1}&=&U-2\mu;&&\bm{u}^1_2&=&\frac{(1,0,-1)^{T}}{\sqrt{2}},\\
\varepsilon_{2}^{2}&=&\frac{U+U_{\uparrow\downarrow}}{2}+\sqrt{\left(\frac{U-U_{\uparrow\downarrow}}{2}\right)^2+4\Omega^2}-2\mu;&&\bm{u}^2_2&=&\frac{(-2\sqrt{2}\Omega,U-U_{\uparrow\downarrow}-\varepsilon_{2}^{2}+\varepsilon_{2}^{0},-2\sqrt{2}\Omega)^{T}}{\sqrt{16\Omega^2+(U-U_{\uparrow\downarrow}-\varepsilon_{2}^{2}+\varepsilon_{2}^{0})^2}}.
\end{array}
\end{eqnarray}

The semi-classical description of the MI state, where each site is occupied by $\rho$ bosons, is given by the direct product of the local eigenstates $\otimes_i |\varphi_\rho^0 \rangle_i$ and the energy of the system is $M \varepsilon_\rho^0$, with $M$ being the number of lattice sites. From standard perturbation theory, the second-order correction to the energy due to the perturbation $\sum_i\hat{{h}}^t_i$ is given by
\begin{eqnarray}
\sum_i\sum_{\rho^\prime=\rho\pm 1}\sum_{\lambda^\prime=0}^{\rho^{\prime}} \frac{|\langle \varphi_\rho^0|\hat{{h}}^t_i|\varphi_{\rho^{\prime}}^{\lambda^\prime} \rangle|^2}{\varepsilon_{\rho}^{0}-\varepsilon_{\rho^\prime}^{\lambda^\prime}}
&=&\sum_i\sum_{r=\pm 1} \bm{u}^{0\dagger}_\rho \mathbf{V}_{\rho,i}^{\rho+r \dagger} \mathbf{U}_{\rho+r} \mathbf{G}_{\rho+r} \mathbf{U}_{\rho+r}^\dagger  \mathbf{V}_{\rho,i}^{\rho+r} \bm{u}^0_\rho\nonumber\\
&=&\sum_i \left(\begin{array}{cc}
\sum_j t_{ji}^\uparrow\psi_{j\uparrow}^*  &  \sum_jt_{ji}^\downarrow\psi_{j\downarrow}^* \end{array}\right) \left(\begin{array}{cc}
a_{\uparrow\uparrow} & a_{\uparrow\downarrow}\\
a_{\downarrow\uparrow}&a_{\downarrow\downarrow}\end{array}\right) \left(\begin{array}{c}
\sum_jt_{ij}^\uparrow\psi_{j\uparrow}  \\  \sum_jt_{ij}^\downarrow\psi_{j\downarrow}\end{array}\right), \label{ssecond_order}
\end{eqnarray}
where $\mathbf{V}_{\rho,i}^{\rho^\prime}$ is a $(\rho^\prime+1)\times (\rho+1)$ matrix whose components are $\langle n_1^\prime,n_2^\prime| \hat{{h}}^t_i|n_1,n_2 \rangle$ with $n_1^\prime+n_2^\prime=\rho^\prime$ and $n_1+n_2=\rho$, $\mathbf{U}_{\rho^\prime}$ is a $(\rho^\prime+1)\times (\rho^\prime+1)$ eigenvector matrix matrix with rows defined by $\left(\bm{u}^{0}_{\rho^\prime}~\bm{u}^1_{\rho^\prime}~\cdots~\bm{u}^{\rho^\prime}_{\rho^\prime}\right)$, and $\mathbf{G}_{\rho^\prime}$ is a $(\rho^\prime+1)\times (\rho^\prime+1)$ diagonal matrix with diagonal components $\left\{\frac{1}{\varepsilon_{\rho}^{0}-\varepsilon_{\rho^\prime}^{0}},\frac{1}{\varepsilon_{\rho}^{0}-\varepsilon_{\rho^\prime}^{1}},\cdots,\frac{1}{\varepsilon_{\rho}^{0}-\varepsilon_{\rho^\prime}^{\rho^\prime}}\right\}$. In the first line of Eq.~(\ref{ssecond_order}), we need to consider only the eigenstates with filling factor $\rho^\prime=\rho\pm 1$ as intermediate states, since the perturbation $\sum_i\hat{{h}}^t_i$ creates or annihilates a single boson. The explicit expressions of $\mathbf{V}_{\rho,i}^{\rho\pm 1}$ for $\rho=2$ are given by
\begin{eqnarray}
\mathbf{V}_{2,i}^{1}=\left(\begin{array}{ccc}
-\sqrt{2}\sum_j t_{ji}^\uparrow\psi_{j\uparrow}^* &  -\sum_j t_{ji}^\downarrow\psi_{j\downarrow}^* & 0\\
0&-\sum_j t_{ji}^\uparrow\psi_{j\uparrow}^* &-\sqrt{2}\sum_j t_{ji}^\downarrow\psi_{j\downarrow}^*\end{array}\right)
\end{eqnarray}
and 
\begin{eqnarray}
\mathbf{V}_{2,i}^{3}=\left(\begin{array}{ccc}
-\sqrt{3}\sum_j t_{ij}^\uparrow\psi_{j\uparrow} &0&0\\
 -\sum_j t_{ij}^\downarrow\psi_{j\downarrow} & -\sqrt{2}\sum_j t_{ij}^\uparrow\psi_{j\uparrow}&0\\
0&-\sqrt{2}\sum_j t_{ij}^\downarrow\psi_{j\downarrow}&-\sum_j t_{ij}^\uparrow\psi_{j\uparrow} \\
0&0& -\sqrt{3}\sum_j t_{ij}^\downarrow\psi_{j\downarrow}
\end{array}\right).
\end{eqnarray}
The matrix elements $a_{ss^\prime}$ in the last line of Eq.~(\ref{ssecond_order}) are only a function of the paramters $\mu$, $\Omega$, $U$, $U_{\uparrow\downarrow}$, and $\rho$. 
Therefore, together with the constant term $\sum_{ijs}t_{ij}^s \psi_{is}^{\ast}\psi_{js}$ caused by the Gutzwiller decoupling, the total second-order contribution in $\psi_{is}$ can be written in Fourier space as
\begin{eqnarray}
\sum_{\bm{k}} \left(\begin{array}{cc}
\psi_{\bm{k}\uparrow}^*  &  \psi_{\bm{k}\downarrow}^* \end{array}\right) \left(\begin{array}{cc}
a_{\uparrow\uparrow} \epsilon_{\bm{k}\uparrow}^2-\epsilon_{\bm{k}\uparrow}&  a_{\uparrow\downarrow}\epsilon_{\bm{k}\uparrow}\epsilon_{\bm{k}\downarrow}\\
a_{\downarrow\uparrow}\epsilon_{\bm{k}\downarrow}\epsilon_{\bm{k}\uparrow}&a_{\downarrow\downarrow} \epsilon_{\bm{k}\downarrow}^2-\epsilon_{\bm{k}\downarrow}\end{array}\right) \left(\begin{array}{c}
\psi_{\bm{k}\uparrow}  \\  \psi_{\bm{k}\downarrow}\end{array}\right)=
\sum_{\bm k}\left( \xi({\bm k})\left|\tilde{\psi}_{{\bm k}}^{(-)}\right|^2+\xi^+({\bm k})\left|\tilde{\psi}_{{\bm k}}^{(+)}\right|^2\right)
\label{s2nde}\end{eqnarray}
with $\epsilon_{\bm{k}s}=-2t(\cos (k_x+s{k_T})+\cos k_y+\cos k_z)$. (Note that $a_{\downarrow\uparrow}=a_{\uparrow\downarrow}^*$.) Here, the diagonalization with respect to the pseudospin index was performed by the transformation\begin{eqnarray}
\left(\begin{array}{c}
\psi_{\bm{k}\uparrow}  \\  \psi_{\bm{k}\downarrow}\end{array}\right)=\left(\begin{array}{cc}
\cos \theta_{\bm k}&  -e^{-i\chi}\sin \theta_{\bm k}\\
e^{i\chi}\sin \theta_{\bm k}&\cos \theta_{\bm k}\end{array}\right) \left(\begin{array}{c}
\tilde{\psi}_{\bm{k}}^{(-)}  \\  \tilde{\psi}_{\bm{k}}^{(+)}\end{array}\right),
\end{eqnarray}
which gives the following two branches for the excitation energies above the MI state with filling factor $\rho$:
\begin{eqnarray}
\xi ({\bm k})&=&\frac{a_{\uparrow\uparrow} \epsilon_{\bm{k}\uparrow}^2-\epsilon_{\bm{k}\uparrow}+a_{\downarrow\downarrow} \epsilon_{\bm{k}\downarrow}^2-\epsilon_{\bm{k}\downarrow}}{2}-\sqrt{\left(\frac{a_{\uparrow\uparrow} \epsilon_{\bm{k}\uparrow}^2-\epsilon_{\bm{k}\uparrow}-a_{\downarrow\downarrow} \epsilon_{\bm{k}\downarrow}^2+\epsilon_{\bm{k}\downarrow}}{2}\right)^2+a_{\uparrow\downarrow}a_{\downarrow\uparrow}\epsilon_{\bm{k}\uparrow}^2\epsilon_{\bm{k}\downarrow}^2}\nonumber\\
\xi^+ ({\bm k})&=&\frac{a_{\uparrow\uparrow} \epsilon_{\bm{k}\uparrow}^2-\epsilon_{\bm{k}\uparrow}+a_{\downarrow\downarrow} \epsilon_{\bm{k}\downarrow}^2-\epsilon_{\bm{k}\downarrow}}{2}+\sqrt{\left(\frac{a_{\uparrow\uparrow} \epsilon_{\bm{k}\uparrow}^2-\epsilon_{\bm{k}\uparrow}-a_{\downarrow\downarrow} \epsilon_{\bm{k}\downarrow}^2+\epsilon_{\bm{k}\downarrow}}{2}\right)^2+a_{\uparrow\downarrow}a_{\downarrow\uparrow}\epsilon_{\bm{k}\uparrow}^2\epsilon_{\bm{k}\downarrow}^2}.
\end{eqnarray}
The components of the diagonalization matrix satisfy the conditions
\begin{eqnarray}
\sin 2\theta_{\bm k}=\frac{2\sqrt{a_{\uparrow\downarrow}a_{\downarrow\uparrow}\epsilon_{\bm{k}\uparrow}^2\epsilon_{\bm{k}\downarrow}^2}}{\xi^+ ({\bm k})-\xi ({\bm k})},~\cos 2\theta_{\bm k}=\frac{a_{\uparrow\uparrow} \epsilon_{\bm{k}\uparrow}^2-\epsilon_{\bm{k}\uparrow}-a_{\downarrow\downarrow} \epsilon_{\bm{k}\downarrow}^2+\epsilon_{\bm{k}\downarrow}}{\xi^+ ({\bm k})-\xi({\bm k})},~{\rm and}~\chi=-{\rm Arg} [a_{\uparrow\downarrow}].
\end{eqnarray}
If the lower branch of excitation energies $\xi ({\bm k})$ is positive for any value of ${\bm k}$, the minimization of the second-order energy~(\ref{s2nde}) gives the solution $\left|\tilde{\psi}_{{\bm k}}^{(\pm)}\right|=0$, which means that the system remains in the MI state. When the minimum excitation energy of $\xi ({\bm k})$ becomes negative, BEC occurs and the system undergoes a phase transition to a superfluid phase. When the excitation spectrum $\xi ({\bm k})$ exhibits a double minimum structure at ${\bm k}=\bar{\bm q}$ and $-\bar{\bm q}$ with $\bar{\bm q} \equiv(\bar{q},0,0)$, due to the presence of spin-orbit couplings, the excited particles can be condensed at either or both of the two ${\bm k}$ points. Thus we define the condensate order parameters as $\Phi_{\pm\bar{q}}\equiv \tilde{\psi}_{{\bm k}=\pm\bar{\bm q}}^{(-)} /\sqrt{M}$, that is,
\begin{eqnarray}
\psi_{\bm{k}s}=\sqrt{M}\left(\nu_{\bar{\bm q}}^s \Phi_{\bar{q}}\delta_{{\bm k}=\bar{\bm q}}+\nu_{-\bar{\bm q}}^s\Phi_{-\bar{q}}\delta_{{\bm k}=-\bar{\bm q}}\right)~~{\rm with}~~\nu_{\pm\bar{\bm q}}^s\equiv \left\{ \begin{array}{ll}
\cos \theta_{\pm\bar{\bm q}}  &  (s=\uparrow) \\
e^{i\chi}\sin \theta_{\pm\bar{\bm q}}  & (s=\downarrow)
\end{array} \right. .
\label{sorder_parameter}
\end{eqnarray}

For zero detuning $\delta=0$, the energies of the condensates with two opposite momenta (and any superposition of them) are degenerate: $\xi (\bar{\bm q})=\xi (-\bar{\bm q})\equiv -\bar{\mu}$.
In order to lift the degeneracy, it is required to take into account the fourth-order contribution from $\sum_i\hat{{h}}^t_i$: 
\begin{eqnarray}
&&\sum_i\sum_{r=\pm 1}\left[ \bm{u}^{0\dagger}_\rho\Bigg( 
\mathbf{V}_{\rho,i}^{\rho+r \dagger}\mathcal{G}_{\rho+r} \mathbf{V}_{\rho+r,i}^{\rho+2r \dagger} \mathcal{G}_{\rho+2r} \mathbf{V}_{\rho+r,i}^{\rho+2r}\mathcal{G}_{\rho+r}  \mathbf{V}_{\rho,i}^{\rho+r}
+\sum_{r^\prime=\pm 1}\left[\mathbf{V}_{\rho,i}^{\rho+r^\prime \dagger}\mathcal{G}_{\rho+r^\prime} \mathbf{V}_{\rho+r^\prime,i}^{\rho \dagger} \bar{\mathcal{G}}_{\rho} \mathbf{V}_{\rho+r,i}^{\rho}\mathcal{G}_{\rho+r}  \mathbf{V}_{\rho,i}^{\rho+r}\right]\Bigg) \bm{u}^0_\rho\right.\nonumber\\
&&\left.-\left(\bm{u}^{0\dagger}_\rho\mathbf{V}_{\rho,i}^{\rho+r \dagger}\mathcal{G}^2_{\rho+r} \mathbf{V}_{\rho,i}^{\rho+r}\bm{u}^0_\rho\right)\left(\bm{u}^{0\dagger}_\rho\mathbf{V}_{\rho,i}^{\rho+r \dagger}\mathcal{G}_{\rho+r} \mathbf{V}_{\rho,i}^{\rho+r}\bm{u}^0_\rho\right)\right]\nonumber\\
&=&\sum_i \left(\begin{array}{ccc}
\left(\sum_j t_{ji}^\uparrow\psi_{j\uparrow}^*\right)^2  &\sum_{jl} t_{ji}^\uparrow t_{li}^\downarrow\psi_{j\uparrow}^*\psi_{l\downarrow}^*&  \left(\sum_jt_{ji}^\downarrow\psi_{j\downarrow}^*\right)^2 \end{array}\right) \left(\begin{array}{ccc}
b_{\uparrow\uparrow\uparrow\uparrow} & b_{\uparrow\uparrow\uparrow\downarrow}& b_{\uparrow\uparrow\downarrow\downarrow}\\
b_{\uparrow\downarrow\uparrow\uparrow}&b_{\uparrow\downarrow\uparrow\downarrow}& b_{\uparrow\downarrow\downarrow\downarrow}\\
b_{\downarrow\downarrow\uparrow\uparrow}&b_{\downarrow\downarrow\uparrow\downarrow}& b_{\downarrow\downarrow\downarrow\downarrow}\end{array}\right) \left(\begin{array}{c}
\left(\sum_jt_{ij}^\uparrow\psi_{j\uparrow}\right)^2  \\\sum_{jl}t_{ij}^\uparrow t_{il}^\downarrow\psi_{j\uparrow}\psi_{l\downarrow}\\  \left(\sum_jt_{ij}^\downarrow\psi_{j\downarrow}\right)^2\end{array}\right), \label{sfourth_order}
\end{eqnarray}
where we use the notation $\mathcal{G}_{\rho^\prime}\equiv \mathbf{U}_{\rho^\prime} \mathbf{G}_{\rho^\prime} \mathbf{U}_{\rho^\prime}^\dagger$ and $\bar{\mathcal{G}}_{\rho}\equiv \mathbf{U}_{\rho} \bar{\mathbf{G}}_{\rho} \mathbf{U}_{\rho}^\dagger$ with $\bar{\mathbf{G}}_{\rho}$ being obtained from $\mathbf{G}_{\rho}$ via the replacement of the first diagonal component $\frac{1}{\varepsilon_{\rho}^{0}-\varepsilon_{\rho}^{0}}$ by $0$ to remove the initial state $|\varphi_\rho^0 \rangle$ from the perturbation process. The matrix elements $b_{s_1s_2 s_3s_4}$ in the last line of Eq.~(\ref{sfourth_order}) are again only a function of the paramters $\mu$, $\Omega$, $U$, $U_{\uparrow\downarrow}$, and $\rho$. Performing a Fourier transformation, we rewrite Eq.~(\ref{sfourth_order}) in the form:
\begin{eqnarray}
\frac{1}{M}\sum_{\{s_i\}}\sum_{{\bm k}_1,{\bm k}_2,{\bm k}_3,{\bm k}_4}\!\!\!\!\!\delta^{(2\pi)}_{{\bm k}_1+{\bm k}_2,{\bm k}_3+{\bm k}_4}\Pi_{{\bm k}_1{\bm k}_2;{\bm k}_3{\bm k}_4}^{s_1s_2;s_3s_4} \psi_{\bm{k}_1s_1}^* \psi_{\bm{k}_2s_2}^* \psi_{\bm{k}_3s_3} \psi_{\bm{k}_4s_4}=M\left(\frac{\mathit{\Gamma}_1}{2}\left(|\Phi_{\bar{q}}|^4+|\Phi_{-\bar{q}}|^4\right)+\mathit{\Gamma}_2|\Phi_{\bar{q}}|^2|\Phi_{-\bar{q}}|^2\right),\label{seq36}
\end{eqnarray}
where the sum $\sum_{\{s_i\}}$ runs over the spin indices that span the nine components of $b_{s_1s_2s_3s_4}$ shown in Eq.~(\ref{sfourth_order}) and $\Pi_{{\bm k}_1{\bm k}_2;{\bm k}_3{\bm k}_4}^{s_1s_2;s_3s_4} \equiv b_{s_1s_2s_3s_4}\epsilon_{\bm{k}_1s_1}\epsilon_{\bm{k}_2s_2}\epsilon_{\bm{k}_3s_3}\epsilon_{\bm{k}_4s_4}$. Substituting Eq.~(\ref{sorder_parameter}) on the left hand side of Eq.~(\ref{seq36}), we obtain the effective interactions between the condensed particles with same momentum ($\mathit{\Gamma}_1$) and with opposite momenta ($\mathit{\Gamma}_2$) as 
\begin{eqnarray}
\mathit{\Gamma}_1&=&2\sum_{\{s_i\}}\Pi_{\bar{\bm q}\bar{\bm q};\bar{\bm q}\bar{\bm q}}^{s_1s_2;s_3s_4}\nu_{\bar{\bm q}}^{s_1 *}~\nu_{\bar{\bm q}}^{s_2 *}\nu_{\bar{\bm q}}^{s_3}\nu_{\bar{\bm q}}^{s_4}=2\sum_{\{s_i\}}\Pi_{-\bar{\bm q}-\bar{\bm q};-\bar{\bm q}-\bar{\bm q}}^{s_1s_2;s_3s_4}~\nu_{-\bar{\bm q}}^{s_1 *}\nu_{-\bar{\bm q}}^{s_2 *}\nu_{-\bar{\bm q}}^{s_3}\nu_{-\bar{\bm q}}^{s_4} ,\nonumber\\
\mathit{\Gamma}_2&=&\sum_{\{s_i\}}\Big[ \Pi_{\bar{\bm q}-\bar{\bm q};\bar{\bm q}-\bar{\bm q}}^{s_1s_2;s_3s_4}~\nu_{\bar{\bm q}}^{s_1 *}\nu_{-\bar{\bm q}}^{s_2 *}\nu_{\bar{\bm q}}^{s_3}\nu_{-\bar{\bm q}}^{s_4}+\Pi_{\bar{\bm q}-\bar{\bm q};-\bar{\bm q}\bar{\bm q}}^{s_1s_2;s_3s_4}~\nu_{\bar{\bm q}}^{s_1 *}\nu_{-\bar{\bm q}}^{s_2 *}\nu_{-\bar{\bm q}}^{s_3}\nu_{\bar{\bm q}}^{s_4}\nonumber\\
&&~~~~~~~~~~~~~~~~~~~+\Pi_{-\bar{\bm q}\bar{\bm q};-\bar{\bm q}\bar{\bm q}}^{s_1s_2;s_3s_4}\nu_{-\bar{\bm q}}^{s_1 *}\nu_{\bar{\bm q}}^{s_2 *}~\nu_{-\bar{\bm q}}^{s_3}\nu_{\bar{\bm q}}^{s_4}+\Pi_{-\bar{\bm q}\bar{\bm q};\bar{\bm q}-\bar{\bm q}}^{s_1s_2;s_3s_4}~\nu_{-\bar{\bm q}}^{s_1 *}\nu_{\bar{\bm q}}^{s_2 *}\nu_{\bar{\bm q}}^{s_3}\nu_{-\bar{\bm q}}^{s_4} \Big],
\end{eqnarray}
as shown on the right hand side of Eq.~(\ref{seq36}).

In summary, the effective energy function up to the forth-order in the order parameters $\Phi_{\pm\bar{q}}$ is given by
\begin{eqnarray}
\frac{E_{\rm GL}}{M}=\varepsilon_\rho^0-\bar{\mu}\left(|\Phi_{\bar{q}}|^2+|\Phi_{-\bar{q}}|^2\right)+\frac{\mathit{\Gamma}_1}{2}\left(|\Phi_{\bar{q}}|^4+|\Phi_{-\bar{q}}|^4\right)+\mathit{\Gamma}_2|\Phi_{\bar{q}}|^2|\Phi_{-\bar{q}}|^2.\label{sGL_potential}
\end{eqnarray}
The minimization of $E_{\rm GL}$ with respect to the order parameters yields two types of ground states:
\begin{eqnarray}
{\rm For }~\mathit{\Gamma}_1<\mathit{\Gamma}_2,&&|\Phi_{\bar{q}}|^2=\frac{\bar{\mu}}{\mathit{\Gamma}_1}~{\rm and}~\Phi_{-\bar{q}}=0~({\rm or}~{\rm vice}~{\rm versa}),~\frac{E_{\rm GL}}{M}=\varepsilon_\rho^0-\frac{\bar{\mu}^2}{2\mathit{\Gamma}_1};\nonumber\\
{\rm For }~\mathit{\Gamma}_1>\mathit{\Gamma}_2,&&|\Phi_{\bar{q}}|^2=|\Phi_{-\bar{q}}|^2=\frac{\bar{\mu}}{\mathit{\Gamma}_1+\mathit{\Gamma}_2},~\frac{E_{\rm GL}}{M}=\varepsilon_\rho^0-\frac{\bar{\mu}^2}{\mathit{\Gamma}_1+\mathit{\Gamma}_2}.
\end{eqnarray}
Notice that if $\mathit{\Gamma}_1<0$ for $\mathit{\Gamma}_1<\mathit{\Gamma}_2$ or $\mathit{\Gamma}_1+\mathit{\Gamma}_2<0$ for $\mathit{\Gamma}_1>\mathit{\Gamma}_2$, the condensates on the MI phase have negative compressibility, which means that the transition to a superfluid phase becomes first-order.

For $\mathit{\Gamma}_1<\mathit{\Gamma}_2$ and $\bar{q}\neq 0$, the transition from the MI phase spontaneously breaks the $Z_2$ symmetry regarding $\bar{q}$ or $-\bar{q}$ in addition to breaking the global $U(1)$ gauge symmetry. In this case, the condensation yields the transition from the MI phase to a phase separated (PS) state when the condition $\rho_\uparrow=\rho_\downarrow$ is imposed. For $\mathit{\Gamma}_1>\mathit{\Gamma}_2$, although the amplitudes of the two order parameters are identical ($|\Phi_{\bar{q}}|=|\Phi_{-\bar{q}}|\equiv\Phi\neq 0$), the relative phase of the two condensates $\phi={\rm Arg}(\Phi_{\bar{q}}/\Phi_{-\bar{q}})$ is undetermined as can be seen from Eq.~(\ref{sGL_potential}). When $\bar{q}/\pi$ is an irreducible fraction $\zeta/\eta$ with $\zeta$ and $\eta$ being integers, the $\eta$-particle umklapp scattering process $\mathit{\Gamma}^\prime_\eta ((\Phi_{\bar{q}}^\ast)^\eta(\Phi_{-\bar{q}})^\eta+(\Phi_{-\bar{q}}^\ast)^\eta(\Phi_{\bar{q}})^\eta)=2\mathit{\Gamma}^\prime_\eta \Phi^{2\eta} \cos \eta \phi$ takes place since $\eta \bar{q}-(-\eta \bar{q})=0$ (modulo $2\pi$), which determines the relative phase $\phi$. In this case, the condensation yields the transition from the MI phase to the striped superfluid (ST) or the chiral superfluid (CSF) phases.

Notice that the $\eta$-particle process $\propto \cos \eta \phi$ still possesses $\eta$-fold degeneracy in the determination of the relative phase $\phi$. Therefore, at the transition from the MI to ST (or CSF) phase with $\bar{q}/\pi$ being an irreducible fraction $\zeta/\eta$, the $Z_\eta$ symmetry is spontaneously broken as well as the $U(1)$ gauge symmetry. This corresponds to a discrete translational symmetry that the ST state has both in the amplitude and phase of the order parameter in real space. For the CSF state, the amplitude of the order parameter is uniform in real space, but its phase breaks chiral $Z_2$ symmetry. When $\bar{q}/\pi$ is an irrational number, umklapp processes are absent and as a result the additional phase terms just discussed do not appear in the Gintzburg-Landau energy. Therefore, the relative phase is undetermined and is spontaneously chosen by the system in its ground state. In this case, the MI-ST transition is associated with spontaneous $U(1)\times U(1)$ symmetry breaking, rather than the $U(1) \times Z_\eta$ symmetry breaking that occurs in the commensurate case.

The value of $\mathit{\Gamma}^\prime_\eta$ can be calculated in a similar way to those of $\mathit{\Gamma}_1$ and $\mathit{\Gamma}_2$, and in general it has a very complex structure. In the simplest example, when $\bar{q}=\pi/2$ the coefficient of the two-particle umklapp process $2\mathit{\Gamma}^\prime_{\eta=2} \Phi^4 \cos 2 \phi$ is given by
\begin{eqnarray}
\mathit{\Gamma}^\prime_{\eta=2}&=&\sum_{\{s_i\}}\Pi_{\bar{\bm q}\bar{\bm q};-\bar{\bm q}-\bar{\bm q}}^{s_1s_2;s_3s_4}~\nu_{\bar{\bm q}}^{s_1 *}\nu_{\bar{\bm q}}^{s_2 *}\nu_{-\bar{\bm q}}^{s_3}\nu_{-\bar{\bm q}}^{s_4}=\sum_{\{s_i\}}\Pi_{-\bar{\bm q}-\bar{\bm q};\bar{\bm q}\bar{\bm q}}^{s_1s_2;s_3s_4}~\nu_{-\bar{\bm q}}^{s_1 *}\nu_{-\bar{\bm q}}^{s_2 *}\nu_{\bar{\bm q}}^{s_3}\nu_{\bar{\bm q}}^{s_4}.
\end{eqnarray}
This completes our analysis of the Ginzburg-Landau theory used to describe the  transition between the superfluid and the Mott insulator phases.

\end{document}